\documentclass[letterpaper,9pt,twocolumn]{autart}
\usepackage[]{graphicx}
\usepackage{cite}
\usepackage[dvips]{color}
\usepackage{amsfonts}
\usepackage{amsmath}
\usepackage{dsfont}
\usepackage{amssymb}
\usepackage{multirow}
\usepackage{subfigure}
\usepackage{psfrag}
\usepackage[thinlines]{easymat}
\usepackage[dcucite]{harvard}

\newcommand{\rr}{\mathop{{\rm I}\mskip-4.0mu{\rm R}}\nolimits}

\newcommand{\vet}[1]{\ensuremath{{\mathbf #1}}}
\newtheorem{theorem}{\textbf{Theorem}} 
\newtheorem{proposition}{\textbf{Proposition}} 

\newtheorem{definition}{\textbf{Definition}}
\newtheorem{lemma}{\textbf{Lemma}}

\newtheorem{problem}{\textbf{Problem}}
\newtheorem{remark}{\textbf{Remark}}
\newtheorem{example}{\textbf{Example}}
\newtheorem{property}{\textbf{Property}}

\begin{document}
\runauthor{}
\begin{frontmatter}
\title{The\hspace{0.27cm}Deformed\hspace{0.27cm}Consensus\hspace{0.27cm}Protocol\vspace{0.22cm}\\
\normalsize{Extended Version}}

\thanks[footnoteinfo]{The author is currently with the Networked Controlled System (NeCS)
team, Inria Grenoble Rh\^{o}ne-Alpes, 655 \mbox{Avenue} de l'Europe, Montbonnot,
38334 Saint Ismier cedex, France, telephone ($+$04) 76 61 53 22. 
The~author gratefully acknowledges the sponsoring of this work by the COMET K2 center 
``Austrian Center of Competence in \mbox{Mechatronics}'' (ACCM).}

\author[INRIA]{Fabio Morbidi}\ead{fabio.morbidi@inria.fr}
\address[INRIA]{Institute for Design and Control of Mechatronical Systems,
Johannes Kepler University, \mbox{Altenbergerstra\ss e 69, 4040 Linz, Austria}}


\begin{abstract}
This paper studies a generalization of the standard continuous-time consensus protocol, obtained
by replacing the Laplacian matrix of the communication graph with the so-called
\emph{deformed Laplacian}. The deformed Laplacian is a second-degree matrix polynomial in the
real variable $s$ which reduces to the standard Laplacian for $s$ equal to unity.
The~stability properties of the ensuing \emph{deformed consensus protocol} are studied in terms
of parameter $s$ for some special families of undirected and directed graphs, and for arbitrary graph
topologies by leveraging the spectral theory of quadratic eigenvalue problems. Examples and
simulation results are provided to illustrate our theoretical~findings. 
\end{abstract}
\end{frontmatter}
\thispagestyle{empty} 
\pagestyle{empty} 
\pagestyle{plain}

\section{Introduction}\label{SEC:intro}

In the last decade we have witnessed a spurt of interest in multi-agent systems research, in the
control, telecommunication and robotics communities~\cite{BulloCoMa_book09,MesbahiEg_book10,Zampieri_IFAC08,Springer_Handbook08_ch41}.
Distributed control and consensus problems~\cite{OlfatiFaMu_IEEE07,RenBeAt_CSM07}, 
have had a large share in this research activity. Consensus theory originated from the work
of Tsitsiklis~\cite{Tsitsiklis_PhD84}, Jadbabaie \emph{et al.}~\cite{JadbabaieJiMo_TAC03}
and Olfati-Saber \emph{et al.}~\cite{OlfatiSaberMu_TAC04}, in which the consensus problem
was formulated for the first time in system-theoretical terms.
A very rich literature emanated from these seminal contributions in recent years. In particular,
numerous extensions to the prototypal consensus protocol in~\cite{OlfatiSaberMu_TAC04}
have been proposed: among them, we limit
ourselves to mention here the cases of time-varying network topology~\cite{Moreau_TAC05,RenBe_TAC05},
of networks with delayed~\cite{OlfatiSaberMu_TAC04} or quantized\big/noisy communication and link
failure~\cite{FrascaCaFaZa_IJRNC09,KarMo_TSP09}, of random networks~\cite{PorfiriSt_TAC07,TahbazJa_TAC08,FagnaniZa_JSAC08},
of networks with antagonistic interactions~\cite{Altafini_PLOSONE12,Altafini_TAC13},
of distributed average tracking~\cite{SpanosOlMu_IFAC05,YangFrLy_TAC08,ChenCaRe_TAC12}, 
of finite-time consensus~\cite{WangXi_TAC10,Kibangou_ACC12}, 
of logical~\cite{FagioliniViBi_CDC08} and nonlinear
agreement~\cite{BausoGiPe_SCL06,Cortes_AUTO08}, 
and of consensus over finite fields~\cite{PasqualettiBoBu_AUTO13}.\\
This paper follows this active line of research and proposes an original 
extension to the basic continuous-time consensus protocol in~\cite{OlfatiSaberMu_TAC04},
that exhibits a rich variety of behaviors and whose flexibility makes it ideal for a broad range of mobile
robotic \mbox{applications} (e.g., for clustering, or for containment and formation control).
%
%
The~new protocol, termed \emph{deformed consensus protocol}, relies on the
so-called \emph{deformed Laplacian matrix}, a second-degree matrix polynomial in the real variable $s$,
which extends the standard Laplacian matrix and reduces to it for $s$ equal to unity: the deformed Laplacian is indeed
an instance of a more general theory of deformed differential operators developed in mathematical physics in
the last three decades, cf.~\cite[Ch. 18]{HislopSi_book96}. Parameter $s$ has a dramatic
effect on the stability properties of the deformed consensus protocol, and it can be potentially used
by a supervisor to dynamically modify the \mbox{behavior} of the network and trigger
different desired agents' responses according to time-varying external events. 
The~stability properties of the proposed protocol are studied in terms of parameter $s$ for
some special families of undirected and directed graphs for which the
eigenvalues and eigenvectors of the negated deformed Laplacian can be
computed in closed form. In the case
of directed graphs, it is shown that differently from the standard
consensus algorithm, for some values of $s$ the states of the deformed consensus protocol 
may also experience stable steady-state oscillations.
Our analysis is extended to arbitrary graph topologies by exploiting 
the spectral theory of quadratic eigenvalue problems~\cite{TisseurMe_SIAM01}.
The discrete-time version of our consensus protocol, that involves
the so-called \emph{deformed Perron matrix}, is also briefly
discussed.\\
Beside the aforementioned promising applications, we
believe that the study of the proposed protocol is of value for shedding 
new light on known results~\cite{OlfatiSaberMu_TAC04,RenBe_TAC05}, 
and for gaining a more general perspective on consensus algorithms.\\
A preliminary version of this paper appeared in~\cite{Morbidi_CDC12},
compared to which we present here several new theoretical results as
well as more extensive numerical simulations.\\
The rest of the article is organized as follows. In
Sect.~\ref{Sec:prel} we review some relevant notions of
algebraic graph theory. The~main theoretical results of the paper are presented in Sect.~\ref{SEC:Prob}.
In Sect.~\ref{Sect:ext}, two possible extensions of our results are discussed. Finally, in Sect.~\ref{Sect:simul}, 
the theory is illustrated via numerical simulations, and in Sect.~\ref{Sect:concl} the main contributions of
the paper are summarized and possible future research directions are outlined.

\section{Preliminaries}\label{Sec:prel}
%
In this section, we briefly recall some basic notions of algebraic graph
theory that will be used through the paper.
Let $\mathcal{G} = (V,\,E)$ be an undirected graph\footnote{All graphs in this paper
are finite, and with no self-loops and multiple edges.} where
$V = \{1,\ldots,n\}$ is the set of vertices, and $E$ 
is the set of edges~\cite{GodsilRo_book01}.
\begin{definition}[Adjacency matrix $\vet{A}$]\label{Def1}
The adjacency matrix $\vet{A} = [a_{ij}]$ of graph $\mathcal{G}$ is an $n \times n$
matrix defined~as, 
\begin{equation*}\label{Eq:Ad_mat_gen}
a_{ij} \,=\, \left\{
\begin{array}{ll}
1\; & \text{if}\;\; \{i,\,j\} \in E,\vspace{0.1cm}\\
0\; & \text{otherwise}.
\end{array}
\right. \vspace{-0.45cm}
\end{equation*}
\hfill$\diamond$
\end{definition}
\begin{definition}[Laplacian matrix $\vet{L}$]
The Laplacian matrix of graph $\mathcal{G}$ is an $n \times n$ matrix defined as,
$$
\vet{L} \,=\, \vet{D} - \vet{A},
$$
where $\vet{D} = \text{diag}(\vet{A}\mathds{1})$ is the degree 
matrix\footnote{$\text{diag}(\vet{b})$ is a diagonal matrix with the
elements of the vector $\vet{b} \in \rr^n$ put on its main diagonal.} and $\mathds{1} = \mathds{1}_n$ is a
column vector of $n$ ones.~\hfill$\diamond$
\end{definition}
Note that the Laplacian $\vet{L}$ is a symmetric positive semidefinite matrix~\cite{Mohar_GRCA91}.
\begin{property}[Spectral properties of\, $\vet{L}$]\label{Sp_propL}
Let $\lambda_1(\vet{L}) \leq \lambda_2(\vet{L}) \leq \ldots \leq \lambda_n(\vet{L})$ be the
ordered eigenvalues of the Laplacian $\vet{L}$. Then, we have that~\cite{Abreu_LAA07}:
\begin{enumerate}
\item $\lambda_1(\vet{L}) = 0$ 
with corresponding eigenvector $\mathds{1}$. 
The~algebraic multiplicity of $\lambda_1(\vet{L})$ is equal to the number of connected
components in $\mathcal{G}$.
\item $\lambda_2(\vet{L}) > 0$ if and only if the graph $\mathcal{G}$
  is connected. $\lambda_2(\vet{L})$~is called the
  \emph{algebraic connectivity} or \emph{Fiedler value} of the graph
  $\mathcal{G}$.~\hfill$\diamond$
\end{enumerate}
\end{property}
%
\begin{definition}[Bipartite graph] 
A graph $\mathcal{G}$ is called bipartite if its vertex set $V$ can be divided into
two disjoint sets $V_1$ and $V_2$, such that every edge connects a vertex in
$V_1$ to one in $V_2$. 
Equivalently, we have that a graph is bipartite if and only if it does not contain cycles of odd~length.~\hfill$\diamond$
\end{definition}


\begin{definition}[Signless Laplacian matrix $\vet{Q}$] 
The signless Laplacian matrix of graph $\mathcal{G}$ is defined as \cite{CvetkovicRoSi_LAA07}
$$
\vet{Q} \,=\, \vet{D} + \vet{A}. \vspace{-0.65cm}
$$
\hfill$\diamond$
\end{definition}
Note that as $\vet{L}$, the signless Laplacian $\vet{Q}$ is a
symmetric positive semidefinite matrix (but it is not necessarily \mbox{singular}~\cite{BrouwerHa_book12}).
Indeed, $\vet{Q} \,=\, \vet{R}\vet{R}^T$ where $\vet{R}$ is
the vertex-edge incidence matrix\footnote{The vertex-edge incidence
matrix of a graph $\mathcal{G}$ is the \mbox{0-1} matrix $\vet{R} = [r_{ik}]$, with rows indexed by
the vertices and column indexed by the edges, where $r_{ik} = 1$ when
vertex $i$ is an endpoint of edge $k$.} of~$\mathcal{G}$.

\begin{property}[Spectral properties of\, $\vet{Q}$]\label{Prop_Q}
The signless Laplacian $\vet{Q}$ has the following spectral 
properties:
\begin{enumerate} 
\item 
If $\lambda_1(\vet{L}) \leq \lambda_2(\vet{L}) \leq \ldots \leq \lambda_n(\vet{L})$ and 
$\lambda_1(\vet{Q}) \geq \lambda_2(\vet{Q}) \geq \ldots \geq \lambda_n(\vet{Q})$
are the ordered eigenvalues of the Laplacian and signless Laplacian,
respectively,~then we have that~\cite{CvetkovicRoSi_LAA07}:
$$
\lambda_n(\vet{L}) \,\leq\, \lambda_1(\vet{Q}).
$$
Moreover, if $n \geq 2$ we have that
$$
\lambda_2(\vet{L}) \,\leq\, \lambda_2(\vet{Q}) + 2,\vspace{0.05cm}
$$
with equality if and only if $\mathcal{G}$ is the complete graph~$K_n$~\cite[Th.~3.5]{CvetkovicSi_PIMB09}.

\item If $\mathcal{G}$ is a connected graph with $n$ vertices and $m$ edges, then
$$
\lambda_1(\vet{Q}) \leq \frac{2\,m}{n - 1} + n - 2,
$$ 
with equality if and only if $\mathcal{G}$ is the star graph
$K_{1,\,n-1}$ or the complete graph $K_n$~\cite[Th.~1]{CvetkovicSi_AADM10}.

\item We have that
$$
\textup{trace}(\vet{L}) \,=\, \textup{trace}(\vet{Q}) \,=\ 2\,|E|,
$$ 
where $|E|$ denotes the cardinality of the edge set~$E$~\cite{BrouwerHa_book12}. 

\item A graph $\mathcal{G}$ is \emph{regular} (i.e., each vertex
of $\mathcal{G}$ has the same degree) if and only if its
signless Laplacian has an eigenvector whose components 
are all ones~\cite[Prop.~2.1]{CvetkovicSi_PIMB09}.

\item If $\mathcal{G}$ is a \emph{regular graph} of degree $\kappa$ (i.e., each vertex
of $\mathcal{G}$ has the same degree $\kappa \leq n-1$), then,
$$
p_{\vet{L}}(\lambda) = (-1)^{n}\, p_{\vet{Q}}(2\kappa - \lambda),
$$
where $p_{\vet{L}}(\lambda)$ denotes the characteristic polynomial of the Laplacian $\vet{L}$.
If $\mathcal{G}$ is a \emph{bipartite graph}, then~\cite[Prop.~2.3]{CvetkovicRoSi_LAA07},
$$
p_{\vet{L}}(\lambda) \,=\, p_{\vet{Q}}(\lambda).
$$ 

\item The least eigenvalue of $\vet{Q}$ of a connected graph is equal to
$0$ if and only if the graph is \emph{bipartite}. In this case, $0$ is
a simple eigenvalue~\cite[Prop.~2.1]{CvetkovicRoSi_LAA07}.

\item In any graph, the multiplicity of the eigenvalue 0 of $\vet{Q}$ is equal
to the number of \emph{bipartite components} of the 
graph~$\mathcal{G}$~\cite[Prop.~1.3.9]{BrouwerHa_book12}.

\item Let $\mathcal{G}$ be a \emph{regular bipartite graph} of degree $\kappa$. Then the
spectrum of $\vet{Q}$ is symmetric with respect to the point 
$\kappa$~\cite[Prop.~2.2]{CvetkovicSi_PIMB09}.~\hfill$\diamond$ 
\end{enumerate}
\end{property}

Let $\mathcal{D} = (V,\,E)$ be a \emph{directed graph} (or \emph{digraph}, for short) where
$V = \{1,\ldots,n\}$ is the set of vertices and $E \subseteq V \!\times V$
is the set of edges. In the case of directed graphs, we can define the adjacency
and degree matrix, as,
\begin{equation*}\label{Eq:Ad_mat_gen_dir}
a_{ij} \,=\, \left\{
\begin{array}{ll}
1\; & \text{if}\;\; (j,\,i) \in E,\vspace{0.1cm}\\
0\; & \text{otherwise},
\end{array}
\right.
\end{equation*}
and $\vet{D} = \text{diag}(d_{\text{in}}(1),\ldots,d_{\text{in}}(n))$,
where $d_{\textup{in}}(i)$ denotes the in-degree of vertex $i$ with $i
\in \{1,\ldots,n\}$ (i.e., the number
of directed edges pointing at vertex $i$). With~these definitions in
hand, the in-degree Laplacian $\vet{L}(\mathcal{D})$ and 
in-degree signless Laplacian $\vet{Q}(\mathcal{D})$ of $\mathcal{D}$, can be defined as
in the undirected case\footnote{``Out-degree'' versions of
$\vet{L}(\mathcal{D})$ and $\vet{Q}(\mathcal{D})$ can be similarly
introduced, but they will not be considered in this paper.}. Note that 
$\vet{L}(\mathcal{D})$ and~$\vet{Q}(\mathcal{D})$ are nonsymmetric
matrices, and that all the eigenvalues of $\vet{L}(\mathcal{D})$ have
non-negative real parts (this can be easily proved using
Ger\v{s}gorin's disk theorem~\cite{OlfatiFaMu_IEEE07}). 

The following definitions will be used in
Sect.~\ref{Sect:ext_Direct}. 

\begin{definition}[Bipartite digraph]
A digraph $\mathcal{D} = (V,\,E)$ is called bipartite if its vertex set $V$ can be divided into
two disjoint sets $V_1$ and $V_2$, such that $E \cap (V_1 \times V_1) = \emptyset$ and
$E \cap (V_2 \times V_2) = \emptyset$, where $\emptyset$ denotes the
empty set.~\hfill$\diamond$
\end{definition}
%
%
\begin{definition}[Rooted out-branching]
A digraph $\mathcal{D} = (V,\,E)$ is a rooted out-branching 
if~\cite{MesbahiEg_book10}:
\begin{enumerate}
\item It does not contain a directed cycle;
\item It has a vertex $v_{\textup{R}}$ (root) such that for every
other vertex $v \in V$, there is a directed path from $v_{\textup{R}}$ to $v$.~\hfill$\diamond$
\end{enumerate}
\end{definition}

\begin{definition}[Strongly connected digraph]
A~digraph is strongly connected if, between every pair of
distinct vertices, there is a directed path.~\hfill$\diamond$
\end{definition}

\begin{definition}[Weakly connected digraph]
A digraph is weakly connected if its disoriented version 
(i.e. the graph obtained by replacing all its directed edges with undirected ones), 
is connected.~\hfill$\diamond$
\end{definition}

\begin{definition}[Balanced digraph]
A digraph is called balanced if, for every vertex, the in-degree and out-degree are equal, i.e.,
$d_{\textup{in}}(i)\,=\,d_{\textup{out}}(i)$, for all $i \in \{1,\ldots,n\}$.~\hfill$\diamond$
\end{definition}

\section{Deformed consensus protocol}\label{SEC:Prob}

\subsection{Problem formulation}\label{SEC:Form_Prob}

It is well-known~\cite{OlfatiSaberMu_TAC04}, that if the static undirected communication graph $\mathcal{G}$
is connected, each component of the state vector $\vet{x} \triangleq [x_1,\ldots,x_n]^T \in \rr^n$
of the linear time-invariant system,
\begin{equation}\label{Eq_cons_std}
\dot{\vet{x}}(t) \,=\, - \vet{L}\,\vet{x}(t),
\end{equation}
asymptotically converges to the average of the initial states $x_1(0),\ldots,x_n(0)$,
$$
\lim_{t \rightarrow \infty}\; x_i(t) \,=\, \frac{1}{n}\,\sum_{i\,=\,1}^n\,x_i(0) \,=\, \frac{1}{n}\;\vet{x}_0^T\,\mathds{1},
$$
where $\vet{x}_0 \triangleq [x_1(0),\,\ldots,\,x_n(0)]^T$, i.e., \emph{average consensus} is achieved.
The converge rate of the consensus protocol~(\ref{Eq_cons_std}) is dictated by the algebraic
connectivity $\lambda_2(\vet{L})$.\\
Let us now consider the following generalization of the Laplacian $\vet{L}$.
\begin{definition}[Deformed Laplacian $\boldsymbol{\Delta}(s)$]
The~\emph{deformed Laplacian} of the graph $\mathcal{G}$ is an $n
\times n$ matrix \mbox{defined as},
\begin{equation*}\label{Eq:pol_mat}
\boldsymbol{\Delta}(s) \,=\, (\vet{D} - \vet{I}_{n})\,s^2 \,-\, \vet{A}\,s \,+\, \vet{I}_{n},
\end{equation*}
where $\vet{I}_{n}$ is the $n \times n$ identity matrix, and $s$ is a real parameter.~\hfill$\diamond$
\end{definition} 
Note that $\boldsymbol{\Delta}(s)$ is a symmetric matrix (but \emph{not} positive
semidefinite as $\vet{L}$, in general), and that:
$$
\boldsymbol{\Delta}(1) = \vet{L},\quad\; \boldsymbol{\Delta}(-1) = \vet{Q}.
$$
Since $\boldsymbol{\Delta}(0) = \vet{I}_n$, the deformed
Laplacian is a comonic polynomial matrix~\cite[Sect.~7.2]{GohbergLaRo_Book09}. 

The following lemma shows an interesting connection between the spectrum of
the deformed Laplacian and the spectrum of the corresponding
adjacency matrix, for regular graphs.

\begin{lemma}\label{Lemma1}
Let $\mathcal{G}$ be a regular graph of degree $\kappa$. Then,
$$
\lambda_i(\boldsymbol{\Delta}(s)) \,=\, (\kappa - 1)\,s^2 -
\lambda_i(\vet{A})\,s + 1,\;\;\, i \in \{1,\ldots,n\}.\vspace{-0.2cm}
$$
\hfill$\blacksquare$
\end{lemma}
\vspace{-0.15cm}

Inspired by~(\ref{Eq_cons_std}), we will study the
stability properties of the following linear system,
\begin{equation}\label{Eq_cons_defor}
\dot{\vet{x}}(t) = -\boldsymbol{\Delta}(s)\,\vet{x}(t), 
\end{equation}
in terms of the real parameter $s$, assuming that the graph $\mathcal{G}$ is \emph{connected}.
We will refer to~(\ref{Eq_cons_defor}), as the \emph{deformed consensus protocol}.
\begin{figure}[b!]
       \begin{center}
       \psfrag{z}{\footnotesize{$s$ switches}}
       \includegraphics[width=.82\columnwidth]{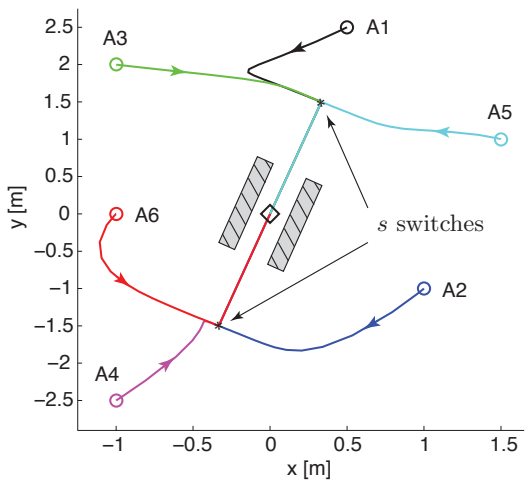}
       \vspace{-0.15cm} 
       \caption{\emph{Illustrative example}: The 6 agents rendezvous
         at the origin while avoiding the two obstacles (gray rectangles): this
         is made possible by switching $s$ from $-1$ to $0$ (the
         initial position of the vehicles is marked with a circle and the final position
         with a diamond).}\label{FIG:Mov_example}
       \end{center}
\end{figure}

\begin{remark}
Note that parameter $s$ in the deformed Laplacian
$\boldsymbol{\Delta}(s)$ can be regarded as a
control input and it can be exploited to \emph{dynamically} modify 
the behavior of system~(\ref{Eq_cons_defor}). This may be useful when the
vertices of the graph are mobile robots and a human supervisor
is interested in changing the collective behavior of the team over
time, cf.~\cite{MilutinovicLi_TRO06}, 
e.g., by switching from a marginally- to an asymptotically-stable
equilibrium point of system~(\ref{Eq_cons_defor}), or between two marginally-stable
equilibria. The former case is illustrated in the example in
Fig.~\ref{FIG:Mov_example}, where the communication graph is the path
graph $P_6$: in order to make the 6 single-integrator agents
rendezvous at the origin while avoiding the
two gray obstacles, the supervisor can initially set $s = -1$ and then
switch to $s = 0$ (cf. Prop.~\ref{Prop_path} in Sect.~\ref{Sect:Spec}).~\hfill$\diamond$
\end{remark} 


Note that since $\boldsymbol{\Delta}(1) = \vet{L}$, we will always achieve
average consensus for $s = 1$. Moreover, since $\boldsymbol{\Delta}(s)$ is real symmetric,
all the eigenvalues of $\boldsymbol{\Delta}(s)$ (which are nonlinear functions of $s$)
are~\emph{real}, and the deformed Laplacian admits
the spectral decomposition $\boldsymbol{\Delta}(s) = \vet{U}(s) \boldsymbol{\Lambda}(s)\vet{U}^T(s)$,
where $\vet{U}(s) = [\vet{u}_1(s)\;\vet{u}_2(s) \,\ldots\, \vet{u}_n(s)]$ $\in \rr^{n \times n}$ is the matrix
consisting of normalized and mutually orthogonal eigenvectors of
$\boldsymbol{\Delta}(s)$ and 
$\boldsymbol{\Lambda}(s) = \text{diag}(\lambda_1(\boldsymbol{\Delta}(s)),\ldots,\lambda_n(\boldsymbol{\Delta}(s)))$.
The solution of~(\ref{Eq_cons_defor}), can thus be written as, 
\begin{equation}\label{Eq_cons_defor_decomp}
\vet{x}(t) \,=\, \sum_{i=1}^n\, \exp(-\lambda_i(\boldsymbol{\Delta}(s))t)\,(\vet{u}_i^T(s)\,\vet{x}_0)\,\vet{u}_i(s).
\end{equation}
In Sect.~\ref{Sect:Spec}, we will focus on some special families of undirected graphs for
which the eigenvalues and eigenvectors of $\boldsymbol{\Delta}(s)$ can be
computed in closed form, and thus the stability properties of system~(\ref{Eq_cons_defor})
can be easily deduced from~(\ref{Eq_cons_defor_decomp}).
In~Sect.~\ref{Sect:Gen}, we will address, instead, the more challenging case
of undirected graphs with arbitrary topology. Finally, some extensions
(the discrete-time case, and the case of directed communication networks), will be discussed in~Sect.~\ref{Sect:ext}. 

\subsection{Stability conditions for special families of graphs}\label{Sect:Spec}

\begin{figure*}[t!]
       \begin{center}
       \begin{tabular}{cccc}
       \psfrag{n}{\hspace{-0.037cm}\scriptsize{$n$}}
       \subfigure[]{\includegraphics[width=.443\columnwidth]{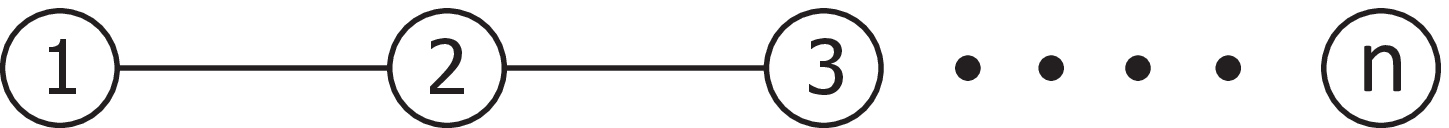}} &
       \psfrag{n}{\hspace{-0.03cm}\footnotesize{$n$}}
       \subfigure[]{\includegraphics[width=.4\columnwidth]{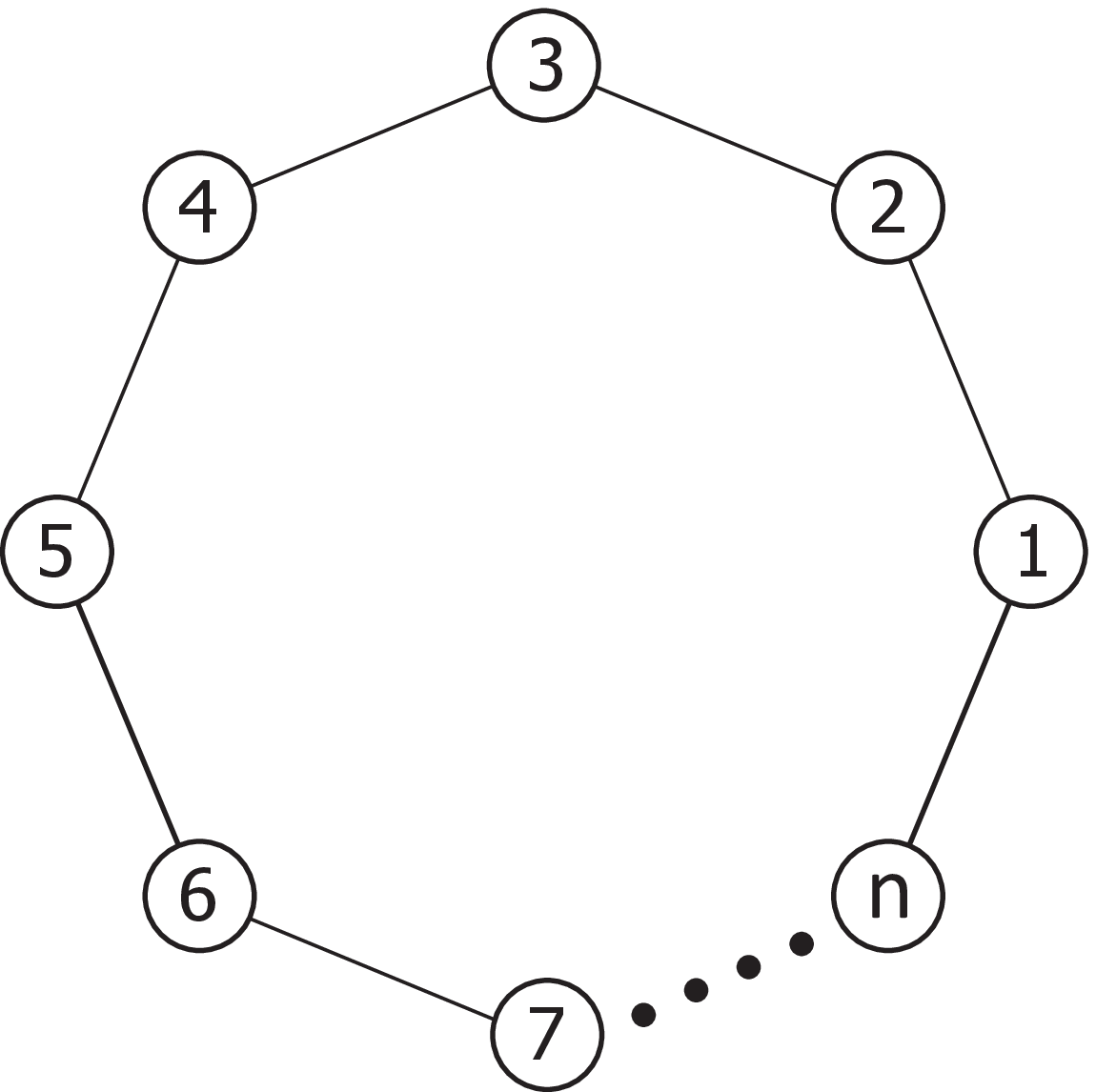}} &
       \subfigure[]{\includegraphics[width=.48\columnwidth]{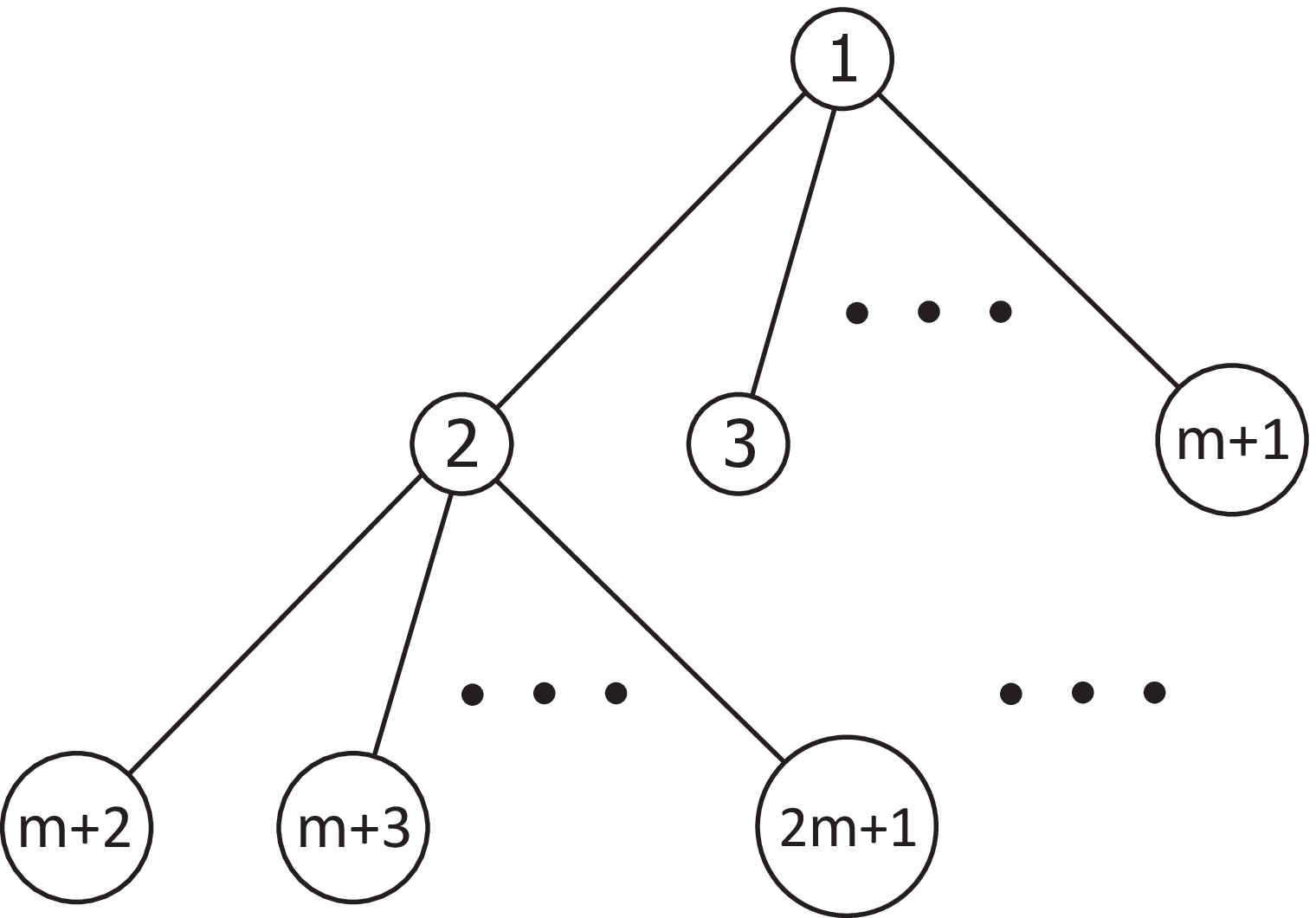}} \;&\;
       \psfrag{n}{\hspace{-0.05cm}\footnotesize{$n$}}
       \subfigure[]{\includegraphics[width=.4\columnwidth]{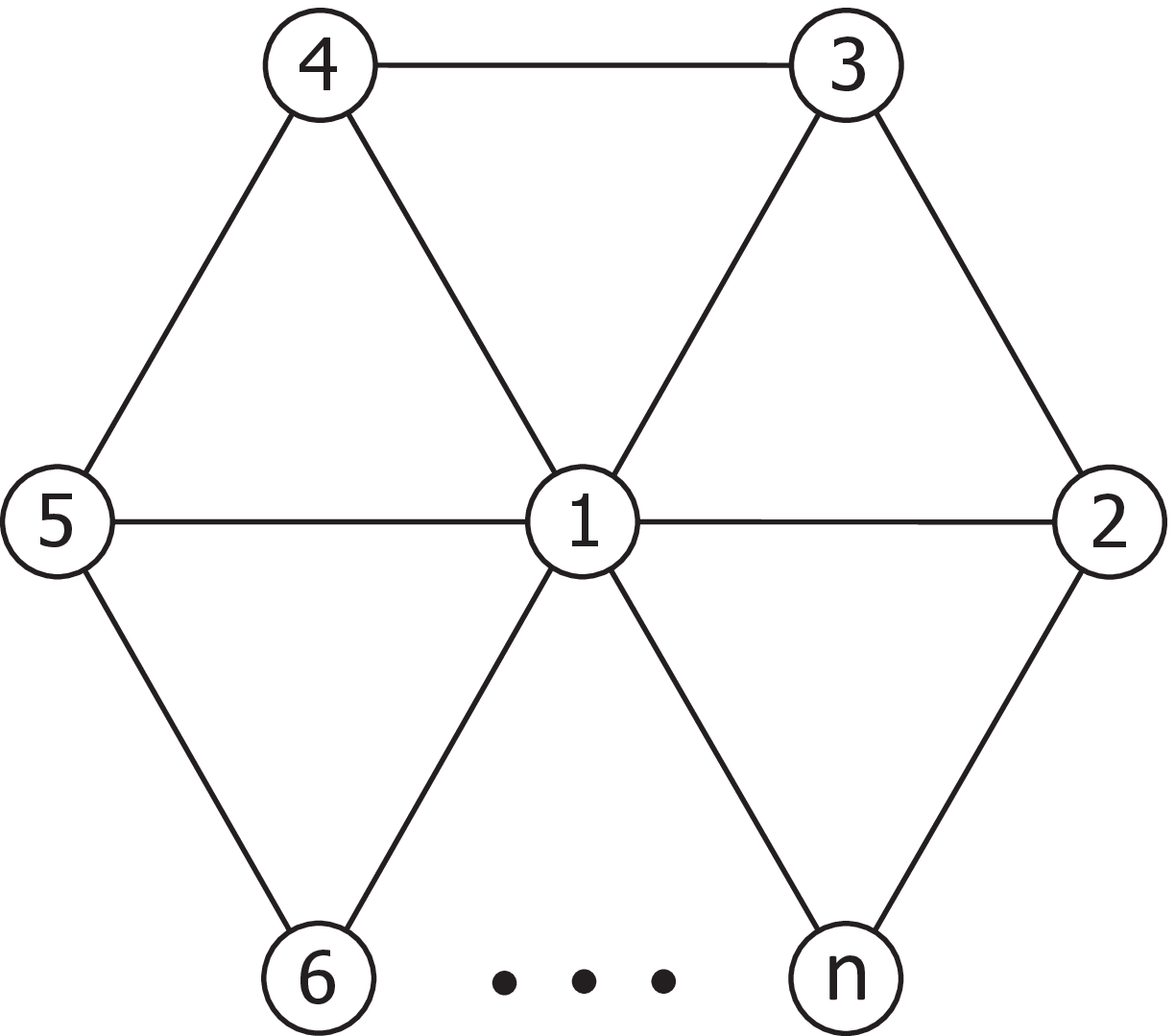}}\vspace{0.1cm}\\
       \end{tabular}
       \begin{tabular}{ccccc}
       \!\!\!\subfigure[]{\includegraphics[width=.457\columnwidth]{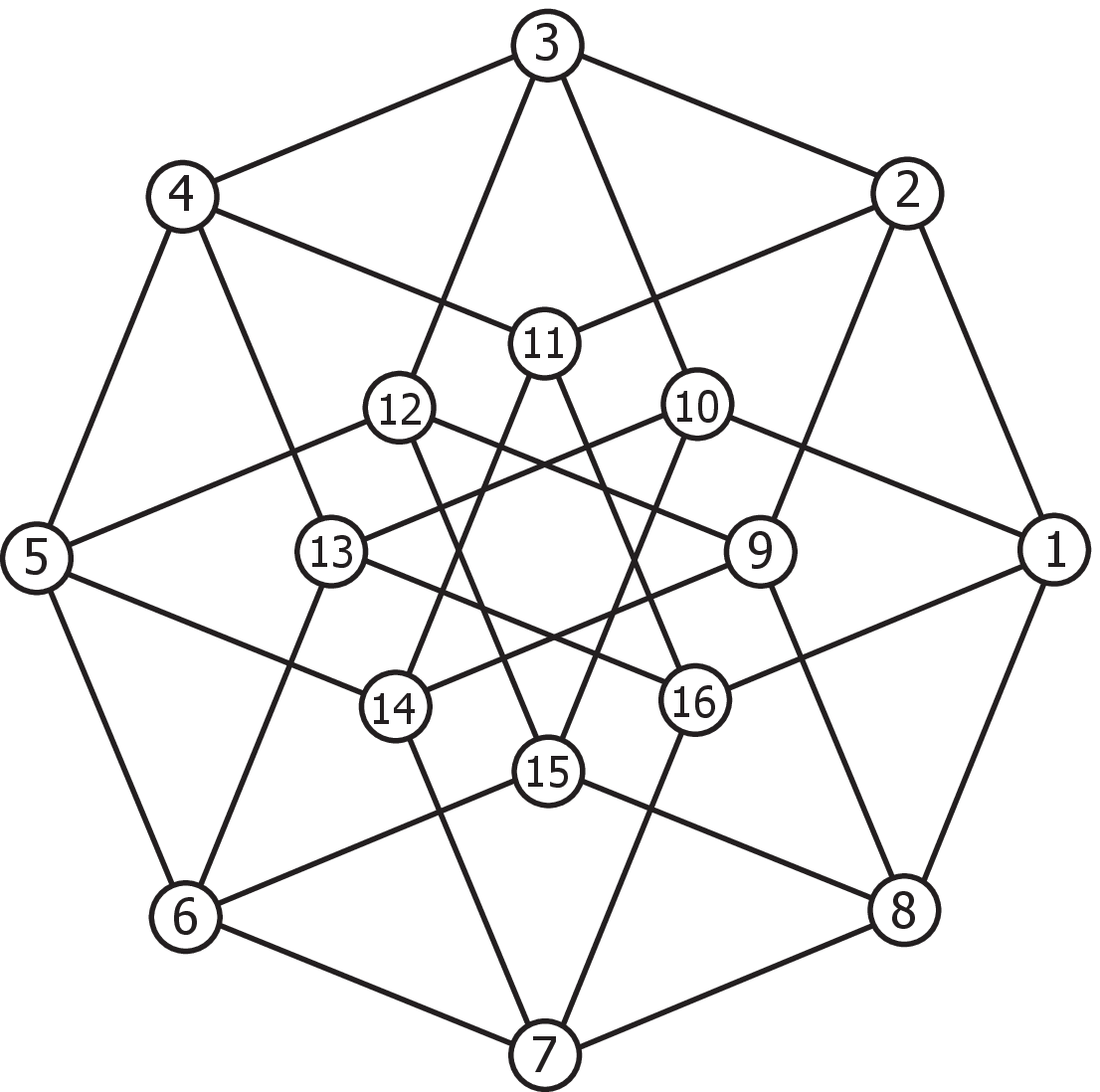}} &
       \subfigure[]{\includegraphics[width=.4\columnwidth]{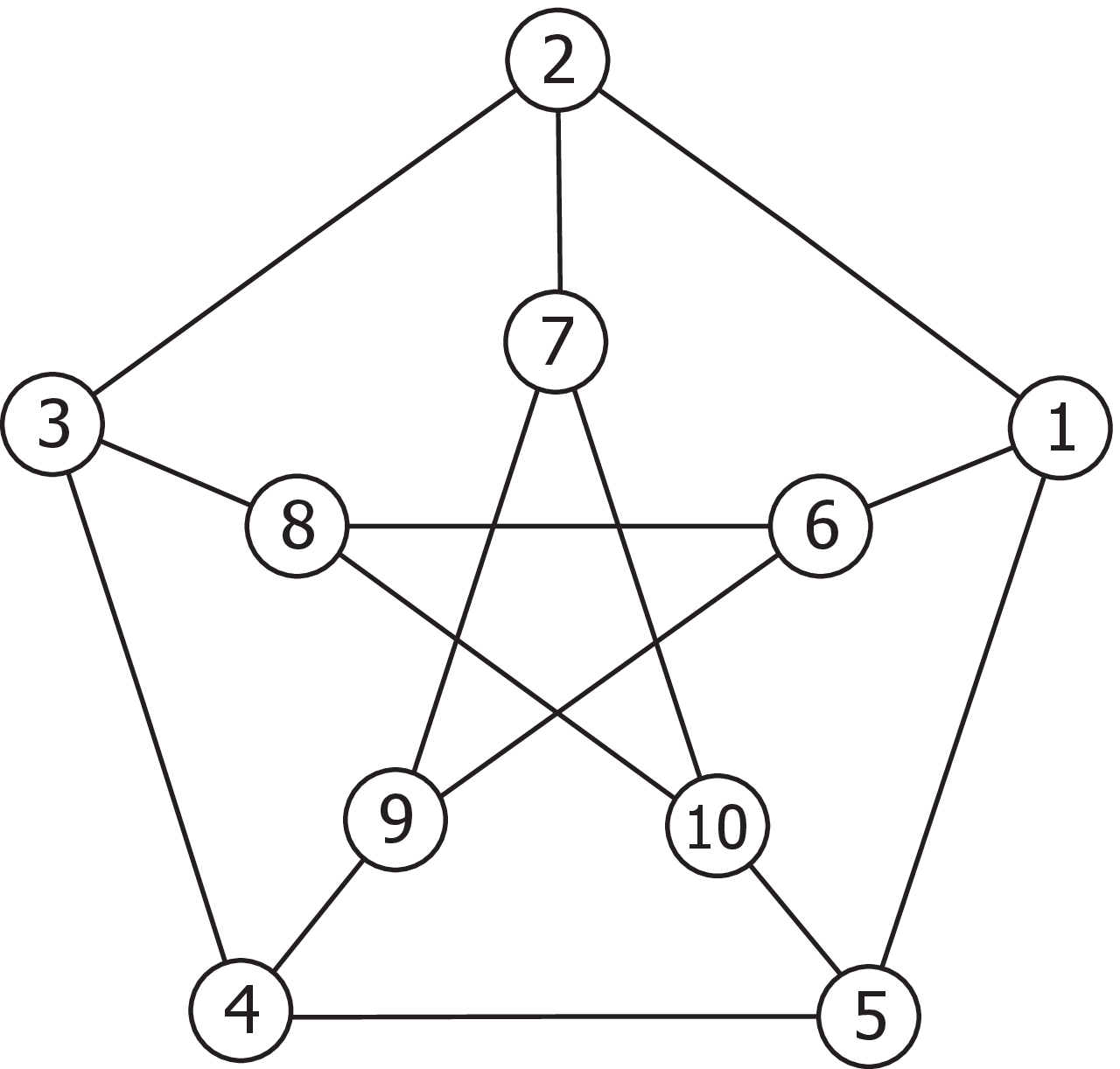}} &
       \psfrag{n}{\hspace{-0.05cm}\footnotesize{$n$}}
       \subfigure[]{\includegraphics[width=.4\columnwidth]{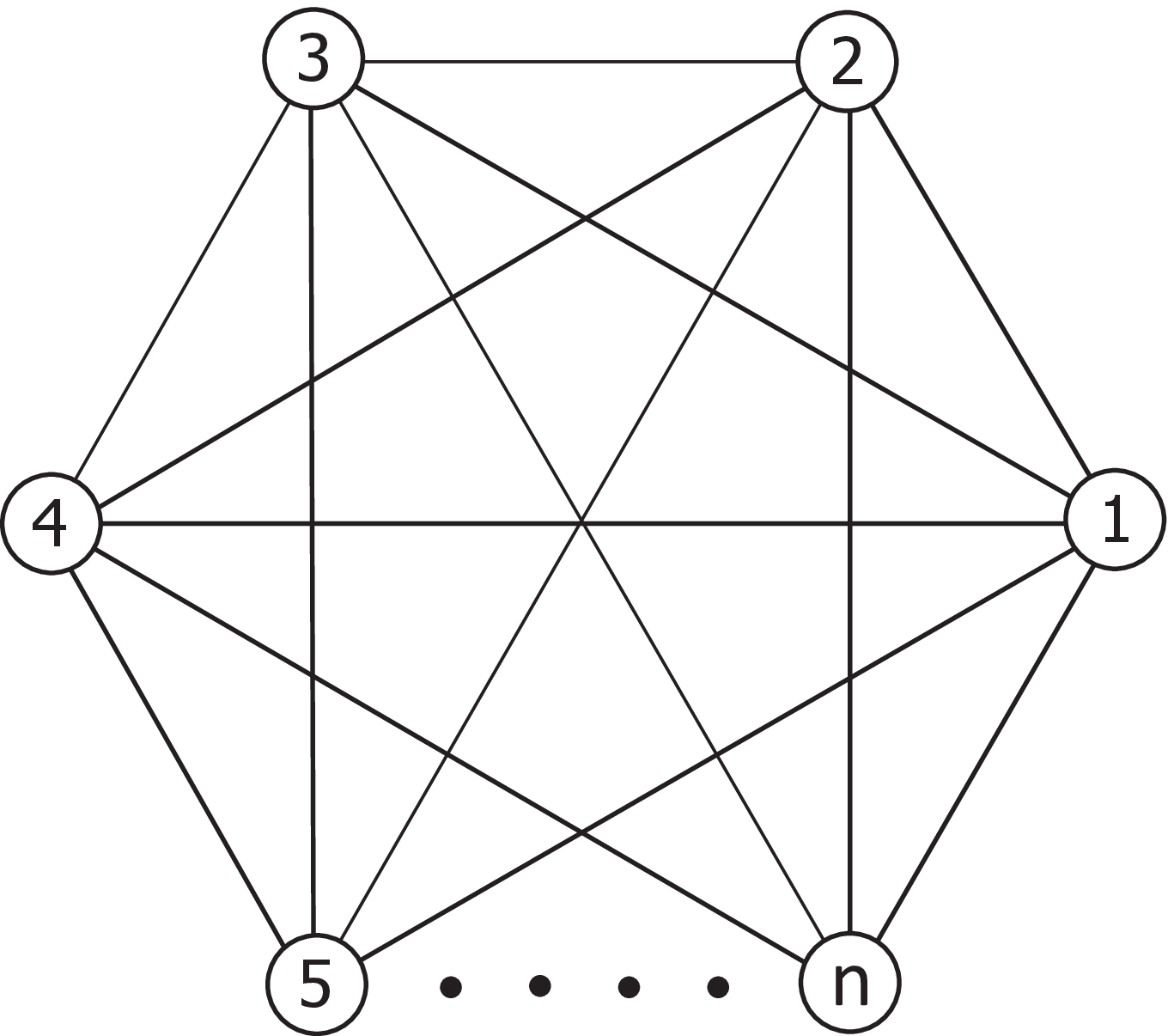}} &
       \psfrag{m}{\hspace{-0.035cm}\scriptsize{$m$}}
       \psfrag{n}{\hspace{-0.038cm}\footnotesize{$n$}}
       \subfigure[]{\includegraphics[width=.285\columnwidth]{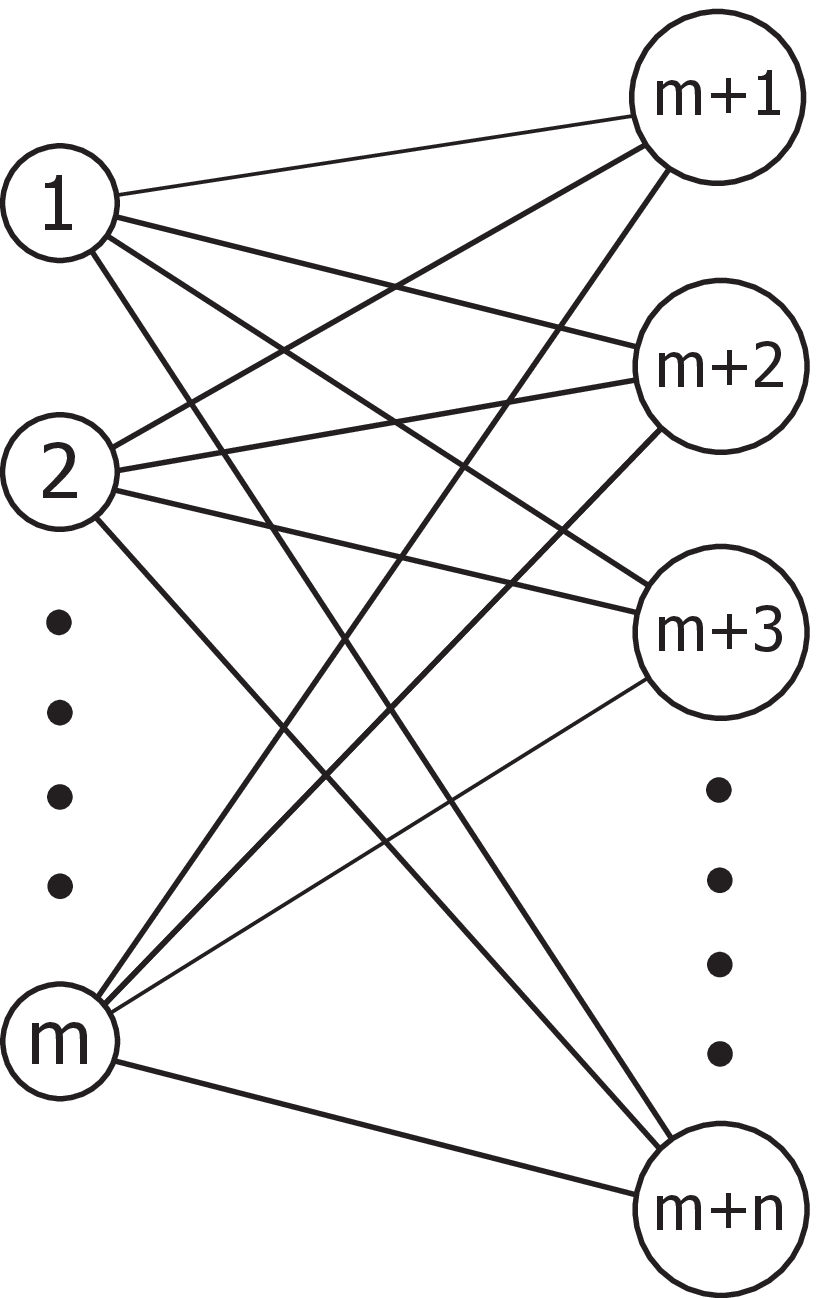}} &
       \psfrag{n}{\hspace{-0.045cm}\footnotesize{$n$}}
       \subfigure[]{\includegraphics[width=.438\columnwidth]{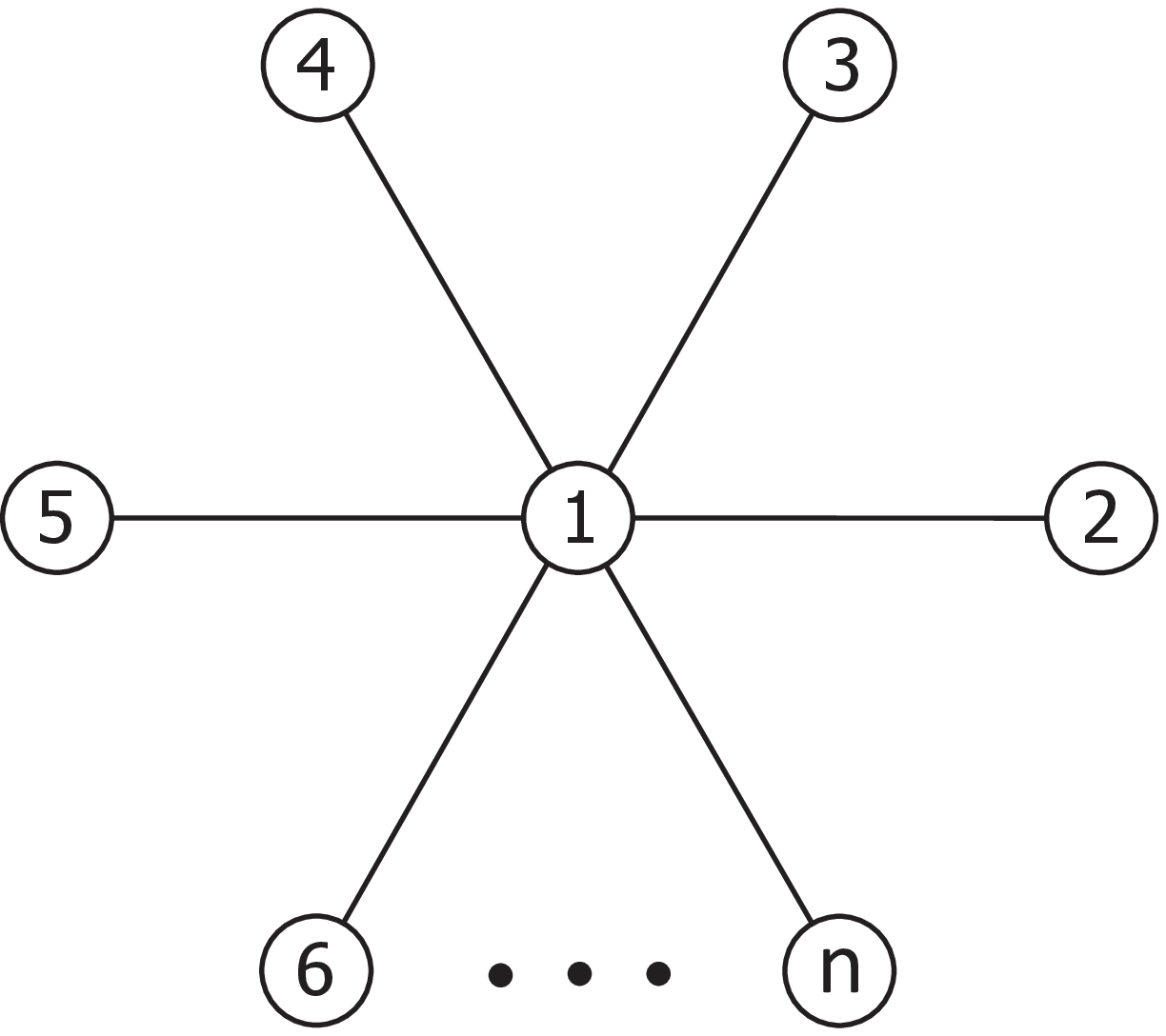}}
       \end{tabular}
           \caption{\emph{Families of undirected graphs}: (a) Path
             graph $P_n$; (b) Cycle graph $C_n$; (c) Full $m$-ary tree; (d) Wheel
             graph $W_n$; (e)~$m$-cube $Q_m$ ($m=4$); (f) Petersen graph;
             (g)~Complete graph $K_n$; (h) Complete bipartite graph
             $K_{m,n}$; (i) Star graph $K_{1,n}$.}
           \label{FIG:Graphs}
       \end{center}
\end{figure*}

This section presents a sequence of nine propositions which provide stability conditions
for system~(\ref{Eq_cons_defor}), in the case of path, cycle, 
full $m$-ary tree, wheel, $m$-cube (or hypercube), Petersen, complete, complete bipartite and star graphs
(see~Fig.~\ref{FIG:Graphs} and refer to~\cite{GodsilRo_book01} for a precise definition of these graphs).
In the following, 
$$
\mathds{k} \,\triangleq\, [-1,\,1,-1,\,1,\ldots,(-1)^{n-1},(-1)^{n}]^T \in \rr^n,
$$ 
$\mathds{J}_{m \times n}$ will denote the $m \times n$ ones matrix, 
$\vet{0}_{m \times n}$ the $m \times n$ zeros matrix, 
$\lfloor \cdot \rfloor$ the floor function which maps a real number to the largest previous integer,
and $\lambda_i(s)$ will be used as a shorthand for 
$\lambda_i(-\boldsymbol{\Delta}(s))$, $i \in \{1,\ldots,n\}$.\\
%
\indent In order to prove our first proposition, we need the following theorem~\cite[Th.~8.5.1]{GolubLo_book96}.
\begin{theorem}[Sturm sequence property]\label{Theo_Sturm}
Consider the following $n \times n$ symmetric tridiagonal matrix,
$$
\vet{T} \,= \left[\,\,
            \begin{matrix}
              a_1 & b_1 & & &\\
              b_1 & a_2 & b_2 & &\vspace{-0.1cm}\\
                & b_2 & a_3 & \!\!\ddots &\vspace{-0.1cm}\\
                & & \!\!\ddots & \ddots & b_{n-1}\\
                & & & b_{n-1} & a_{n}
             \end{matrix}
          \,\right],
$$
where $b_j \neq 0$, $\forall\, j \in \{1,\ldots,n-1\}$. 
Let $\vet{T}^{(r)}$ denote the leading $r \times r$ principal submatrix of $\vet{T}$.
Then, the number of negative eigenvalues of $\vet{T}$
is equal to the number of sign changes in the \emph{Sturm sequence}:
$$
1,\;\;\det(\vet{T}^{(1)}),\;\det(\vet{T}^{(2)}),\,\ldots,\;\det(\vet{T}^{(n)}). 
$$
The result is still valid if zero determinants are encountered
along the way, as long as we define a ``sign change'' to mean a transition from $+$ or 0 to $-$,
or from $-$ or 0 to $+$, but not from $+$ or $-$ to~0.~\hfill$\blacksquare$
\end{theorem}

\begin{proposition}[Path graph $P_n$]\label{Prop_path}
For the path graph $P_n$ with $n \geq 2$ vertices (see Fig.~\ref{FIG:Graphs}(a)), we have that:
\begin{itemize}
\item For $|s| < 1$, system (\ref{Eq_cons_defor}) is asymptotically stable.
\item For $|s| > 1$, system (\ref{Eq_cons_defor}) is unstable.
\item For $s = -1$, system (\ref{Eq_cons_defor}) is marginally stable.
In this case, it is possible to identify two groups of $n/2$ vertices
(if $n$ is even), or one group of $\lfloor n/2 \rfloor$ vertices
and one of $\lfloor n/2 \rfloor + 1$ vertices (if $n$ is odd).
The states associated to the vertices in one group asymptotically
converge to $\frac{1}{n}\,\vet{x}_0^T\mathds{k}$
and the states associated to the vertices in the other group
converge to $-\frac{1}{n}\,\vet{x}_0^T \mathds{k}$.
\end{itemize}
\emph{Proof}:
In this case, $-\boldsymbol{\Delta}(s)$ is a symmetric tridiagonal matrix,
$$
-\boldsymbol{\Delta}(s) = \left[\,
  \begin{matrix}
    -1 & s & & &\\ 
     s & -(s^2 + 1) & s & &\vspace{-0.12cm}\\
     & & \ddots & &\vspace{-0.04cm}\\ 
     & & s & -(s^2 + 1) & s\\
     & & & s & -1
  \end{matrix}
\,\right].
$$
The Sturm sequence of $-\boldsymbol{\Delta}(s)$ is given by
$$
\begin{array}{ll}
1,\;-1,\;1,\;-1,\;1,\,\ldots,\,-1,\;1,\;s^2 - 1, & \text{if}\;\, n\;\, \text{is odd},\vspace{0.15cm}\\
1,\;-1,\;1,\;-1,\;1,\,\ldots,\,1,\;-1,\;1 - s^2, & \text{if}\;\, n\;\, \text{is even}.
\end{array}
$$
Therefore by Theorem~\ref{Theo_Sturm}, if $|s| < 1$, $s \neq 0$,
all the eigenvalues of $-\boldsymbol{\Delta}(s)$ are strictly negative and
system (\ref{Eq_cons_defor}) is asymptotically stable. On the other hand,
the Sturm sequence of $\boldsymbol{\Delta}(s)$, for all $n$, is given by $1,\;1,\ldots,\;1,\;1 - s^2$,
from which we deduce that for $|s| > 1$, $\boldsymbol{\Delta}(s)$ has a negative eigenvalue, and
hence system~(\ref{Eq_cons_defor}) is unstable.
Since $\boldsymbol{\Delta}(0) = \vet{I}_n$, the system
is asymptotically stable for $s = 0$. Finally, for $s = -1$, system (\ref{Eq_cons_defor})
is marginally stable and the unit-norm
eigenvector associated to the zero eigenvalue of $-\boldsymbol{\Delta}(-1)$ is $\frac{1}{\sqrt{n}}\,\mathds{k}$.
\hfill$\blacksquare$
\end{proposition}
\begin{proposition}[Cycle graph $C_n$]\label{Prop_Cycle}
For the cycle graph $C_n$ with $n > 2$ vertices (see Fig.~\ref{FIG:Graphs}(b)), we have that:
\begin{itemize}
\item If $n$ is even:
\begin{itemize}
\item For all $s \in \rr \setminus \{-1,\,1\}$, system (\ref{Eq_cons_defor}) is asymptotically stable.
\item For $s = -1$, system (\ref{Eq_cons_defor}) is marginally stable.
In this case, 
the states associated to $n/2$ vertices asymptotically
converge to $\frac{1}{n}\,\vet{x}_0^T\,\mathds{k}$
and the states associated to the other $n/2$ vertices 
converge to $-\frac{1}{n}\,\vet{x}_0^T\,\mathds{k}$.
\end{itemize}
\item If $n$ is odd, system (\ref{Eq_cons_defor}) is asymptotically stable for all $s \in \rr \setminus \{\,1\}$.
\end{itemize}
\emph{Proof}:
In this case $-\boldsymbol{\Delta}(s)$ is a \emph{circulant matrix}, 
$$
-\boldsymbol{\Delta}(s) = \textup{circ}[-(s^2 +1),\,s,\,0,\,\ldots,\,0,\,s],
$$
i.e., each subsequent row is simply the row above shifted one element
to the right (and wrapped around, i.e., modulo~$n$). The entire matrix is thus determined by the first row.
It is well-known that circulant matrices are diagonalizable by the Fourier matrix $\vet{F}_n$, given via,
$$
\vet{F}^{\ast}_n = \frac{1}{\sqrt{n}}\left[
  \begin{array}{ccccc}
    1 & 1 & 1 & \cdots & 1 \\
    1 & \omega & \omega^2 & \cdots & \omega^{n-1} \\
    1 & \omega^2 & \omega^4 & \cdots & \omega^{2(n-1)} \\
    \vdots & \vdots & \vdots & \ddots & \vdots\vspace{0.1cm}\\
    1 & \omega^{n-1} & \omega^{2(n-1)} & \cdots & \omega^{(n-1)(n-1)}\\
  \end{array}
\!\!\right],
$$
where $\omega \triangleq e^{2\pi j/n}$, $j = \sqrt{-1}$, and $\vet{F}^{\ast}_n$
denotes the conjugate transpose of $\vet{F}_n$,
hence their eigenvalues can be computed in closed form.
The eigenvalues of a general $n \times n$ circulant matrix
$\vet{C} = \text{circ}[c_1,\,c_2,\ldots,\,c_n]$, in fact,
are given~by:
\begin{equation}\label{Eq:Eig_circ}
\lambda_i(\vet{C}) = \rho_{\vet{C}}(\omega^{i-1}),\;\; i \in \{1,\ldots,\,n\},
\end{equation}
where the polynomial $\rho_{\vet{C}}(\xi) = c_n \xi^{n-1} + \ldots + c_3 \xi^{2} + c_2 \xi +~c_1$
is called the circulant's representer~\cite[Th. 3.2.2]{Davis_book94}.
By applying this result to matrix $-\boldsymbol{\Delta}(s)$, for $i \in \{1,\ldots,\,n\}$ we have that,
\begin{equation}\label{Eq:parabola}
\lambda_i(-\boldsymbol{\Delta}(s)) \,=\, -\,s^2 \,+\, 2\,\cos\!\left(\frac{2\,\pi(i-1)}{n}\right)s \,-\,1.
\end{equation}
Observe now that the coordinates of the vertex of the
parabola (\ref{Eq:parabola}) are
$$
\left[\cos\!\left(\frac{2\pi(i - 1)}{n}\right),\,-\sin^2\!\left(\frac{2\pi(i-1)}{n}\right)\!\right],\, i \in \{1,\,\ldots,\,n\}.
$$
If $n$ is even, then $\lambda_i(-\boldsymbol{\Delta}(s)) < 0$, $\forall\, s \in \rr$
and $\forall\, i \neq \{1,\,n/2\,+\,1\}$. For $i = 1$, $\lambda_1(-\boldsymbol{\Delta}(s)) \leq 0$
and $\lambda_1(-\boldsymbol{\Delta}(s)) \,\,{=}\,\, 0$ only for $s = 1$.
For $i = n/2 + 1$, $\lambda_{n/2\,+\,1}(-\boldsymbol{\Delta}(s)) \leq 0$
and $\lambda_{n/2\,+\,1}(-\boldsymbol{\Delta}(s)) = 0$ only for $s = -1$.
The unit-norm eigenvector associated to $\lambda_{n/2\,+\,1}(-\boldsymbol{\Delta}(-1))$ is
$\frac{1}{\sqrt{n}}\,\mathds{k}$.
On the other hand, if $n$ is odd, then $\lambda_i(-\boldsymbol{\Delta}(s)) < 0$,
$\forall\, s \in \rr$ and $\forall\, i \neq 1$. For $i = 1$,
$\lambda_1(-\boldsymbol{\Delta}(s)) \leq 0$ and $\lambda_1(-\boldsymbol{\Delta}(s)) = 0$ only for $s = 1$.
\hfill$\blacksquare$
\end{proposition}
Note that for the cycle graph $C_n$,
$$
\det(-\boldsymbol{\Delta}(s)) \,=\, (-1)^{n}\,(s^{2n} - 2s^n + 1),
$$
and that the $2n$ roots of $\det(-\boldsymbol{\Delta}(s))$ are evenly spaced on the
unit~circle. 

A full \emph{$m$-ary tree} is a rooted tree in which every vertex other than the leaves
has $m$ children ($2$-ary and $3$-ary trees are sometimes called binary and ternary 
trees, respectively~\cite{GodsilRo_book01}).
The \emph{depth} $\delta$ of a vertex is the length of the path from the root
to the vertex. The set of all vertices at a given depth is called a \emph{level} of the tree:
by definition, the root vertex is at depth zero. The number of vertices of a full $m$-ary
tree is $n = \sum_{i=0}^{\delta} m^i$.\vspace{-0.12cm}
\begin{proposition}[Full $m$-ary tree]
For the full $m$-ary tree, with $m \geq 2$ (see Fig.~\ref{FIG:Graphs}(c)), we have that:
\begin{itemize}
\item For $|s| < 1$, system (\ref{Eq_cons_defor}) is asymptotically stable.
\item For $|s| > 1$, system (\ref{Eq_cons_defor}) is unstable.
\item For $s = -1$, system (\ref{Eq_cons_defor}) is marginally stable.
In this case, the states associated to the vertices in the even levels 
of the tree asymptotically converge to $\frac{1}{n}\,\vet{x}_0^T\vet{h}$
while the states associated to the vertices in the odd levels 
of the tree converge to $-\frac{1}{n}\,\vet{x}_0^T\vet{h}$, where, 
$$
\vet{h} \,\triangleq\, [-1,\,\mathds{1}^T_m,\,-\mathds{1}^T_{m^2},\,\mathds{1}^T_{m^3},\,
-\mathds{1}^T_{m^4},\,\mathds{1}^T_{m^5},\,\ldots\,]^T \in \rr^n.
$$
\end{itemize}
\emph{Proof}:
The stability properties of~(\ref{Eq_cons_defor}) are determined in this case by only
one of the eigenvalues of $-\boldsymbol{\Delta}(s)$ (in fact, the other $n-1$ are
negative for all $s$: note that $\det(-\boldsymbol{\Delta}(s)) = s^2-1$).
This eigenvalue is negative for $|s| < 1$, positive for $|s| > 1$ and
zero for $s \in \{-1,\,+1\}$. For $s = -1$, note that,
$$
\qquad\quad\;\lim_{t \,\rightarrow \,
  \infty}\;\exp(-\boldsymbol{\Delta}(-1)\,t)\,\vet{x}_0 \,=\,
\frac{1}{n}\,\vet{h}\,\vet{h}^T\vet{x}_0.\qquad\quad\; 
\hfill\blacksquare
$$
\end{proposition}

\begin{figure}[t!]
  \psfrag{a}{$n$}
  \psfrag{b}{\!$\mu$}
       \begin{center}
           \includegraphics[width=.78\columnwidth]{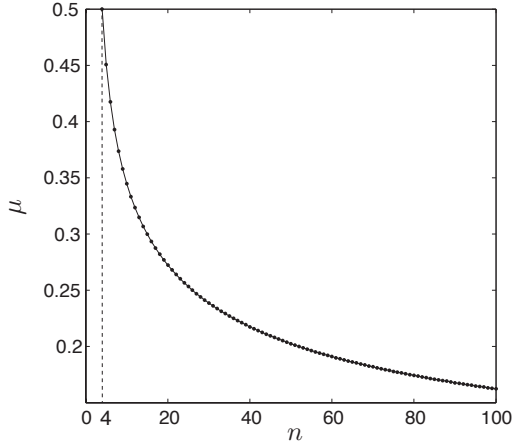}
           \caption{Value of the non-unitary root of~(\ref{Eq:root_mu}) for $n \in [4,\,100]$.}
           \label{FIG:Root}
       \end{center}
\end{figure}

\begin{proposition}[Wheel graph $W_n$]
Consider a wheel graph $W_{n}$ with $n > 3$ vertices where vertex 1 is the center of the wheel (see Fig.~\ref{FIG:Graphs}(d)),
and let $\mu$ be the non-unitary root~of\vspace{-0.15cm}
\begin{equation}\label{Eq:root_mu}
\!-\frac{n}{2}\,s^2 \,+\, s \,+\, \frac{\sqrt{\big((n-4)\,s +
    2\big)^2 + 4(n-1)}}{2}\,\,s \,-\, 1. 
\end{equation}
$\mu$ monotonically decreases from $1/2$ (for $n = 4$) to $0$ (for~$n = \infty$) [see Fig.~\ref{FIG:Root}].
We have that:
\begin{itemize}
\item For $s > 1$ or $s < \mu$, system (\ref{Eq_cons_defor}) is asymptotically stable.
\item For $s \in (\mu,\,1)$, system (\ref{Eq_cons_defor}) is unstable.
\item For $s = \mu$, system~(\ref{Eq_cons_defor}) is marginally
stable. If $n = 4$ average consensus is achieved. Instead, if $n > 4$
the state associated to vertex 1 asymptotically converges to
$\vet{x}^T_0[\alpha^2,\,\alpha\beta,\,\ldots,\alpha \beta]^T$, and the
states associated to the other $n-1$ vertices converge 
to $\vet{x}^T_0[\alpha\beta,\,\beta^2,\,\ldots,\beta^2]^T$,
where $[\alpha,\,\beta,\,\ldots,\beta]^T$, $\alpha,\,\beta \in \rr$, is
the unit-norm eigenvector associated to the zero eigenvalue of~$-\boldsymbol{\Delta}(\mu)$.
\end{itemize}
\emph{Proof}:
The eigenvalues of matrix $-\boldsymbol{\Delta}(s)$ are:
$$
\begin{array}{l}
\!\!\displaystyle\lambda_{1}(s) = -\frac{n}{2}\,s^2 \!+ s + \!\frac{\sqrt{\big((n-4)\,s +
    2\big)^2 + 4(n-1)}}{2}\,s - \!1,\vspace{0.16cm}\\
\!\!\displaystyle\lambda_{2}(s) = -\frac{n}{2}\,s^2 \!+ s - \!\frac{\sqrt{\big((n-4)\,s +
    2\big)^2 + 4(n-1)}}{2}\,s - \!1,\vspace{0.16cm}\\
\!\!\lambda_{i+1}(s) = -2s^2 + 2\cos\!\big(\frac{2\pi(i-1)}{n-1}\big)s - 1,\, i \in \!\{2,\ldots,n-1\}.
\end{array}
$$
Note that the coordinates of the vertex of the
parabola $\lambda_{i+1}(s)$, $i \in \!\{2,\ldots,n-1\}$, are,
$$
\left[\frac{1}{2}\cos\!\Big(\frac{2\,\pi (i-1)}{n-1}\Big),\;-\frac{1}{2}\Big(1 +\sin^2\!\Big(\frac{2\,\pi (i-1)}{n-1}\,\Big)\Big)\right],
$$
therefore $\lambda_{i+1}(s) < 0$, $\forall\, s \in \rr$, $\forall\, i \in \!\{2,\ldots,n-1\}$.
We also have that $\lambda_{2}(s) < 0$, $\forall\, s \in
\rr$. Finally, it is easy to verify that $\lambda_{1}(s)$ has always two roots,
$s = \mu$ and $s = 1$. $\lambda_{1}(s) > 0$ for $s \in (\mu,\,1)$ and $\lambda_{1}(s) < 0$
for $s > 1$ or~$s < \mu$.~\hfill$\blacksquare$
%
\end{proposition}
\begin{table*}[t!]
\renewcommand{\arraystretch}{1.42}
\begin{center}
  \begin{tabular}{|l|c|c|} 
    \hline
    \textbf{Graph name} & \textbf{Asymptotic stability for\,:}
    & \textbf{Marginal stability for\,:}\\
    \hline\hline
    Path graph $P_n$, $n \geq 2$ & $|s| < 1$ & $s = -1$\; (2 groups of vertices)\vspace{0.01cm}\\
    \hline
    Cycle graph $C_n$, $n > 2$, $n$ even & $\forall\, s \in\,
    \rr\setminus \{-1,\,1\}$ &\hspace{0.01cm} $s = -1$\; (2 groups of vertices)\vspace{0.01cm}\\
    \hline
    Cycle graph $C_n$, $n > 2$, $n$ odd & $\forall\, s \in\, \rr\setminus \{1\}$ & \vspace{0.01cm}\\
    \hline
    Full $m$-ary tree, $m \geq 2$ & $|s| < 1$ & $s = -1$\; (2 groups of vertices)\vspace{0.01cm}\\
    \hline
    Wheel graph $W_n$, $n > 3$ &  $s > 1$ \,or\, $s < \mu$ & $s =
    \mu$\, (2 groups of vertices for $n > 4$)\vspace{0.01cm}\\
    \hline
    \multirow{2}{*}{$m$-cube, $Q_m$, $n \,=\, 2^m > 4$} &  \multirow{2}{*}{$|s| > 1$ or $|s| < \frac{1}{m-1}$}
    & $s \in \{-1, -\frac{1}{m-1}\}$\, (2 groups of vertices)\vspace{0.01cm}\\
    & & $s = \frac{1}{m-1}$\; (average consensus)\vspace{0.01cm}\\
    \hline
    Petersen graph & $s > 1$ \,or\, $s < 1/2$ & $s = 1/2$\; (average consensus) \vspace{0.01cm}\\
    \hline
    Complete graph $K_n$, $n > 2$ & $s > 1$ \,or\, $s < \frac{1}{n-2}$ & $s = \frac{1}{n-2}$\; (average consensus) \vspace{0.01cm}\\
    \hline
    \multirow{3}{*}{
    $
    \begin{array}{c}
    \text{Complete bipartite graph}\\
    K_{m,n},\, m,n \geq 2
    \end{array}
    $ 
    }   
    & \multirow{4}{*}{$|s| > 1$ or $|s| < \frac{1}{\sqrt{(m-1)(n-1)}}$} &
    $s \in \{-1,\,- \frac{1}{\sqrt{(m-1)(n-1)}}\}$\;
    (2 groups of vertices)\\
    & & $s = \frac{1}{\sqrt{(m-1)(n-1)}}$\!
    $
    \begin{array}{l}
    (\text{if}\;\, m = n,\;\, \text{average consensus})\\
    (\text{if}\;\, m \neq n,\;\, \text{2 groups of vertices})
    \end{array}
    $
    \vspace{0.01cm}\\\hline
    Star graph $K_{1,n}$, $n \geq 3$ & $|s| < 1$ & $s = -1$\; (2 groups of vertices)\\
    \hline
  \end{tabular}
  \vspace{0.3cm}
  \caption{Summary of the stability properties of the deformed consensus protocol~(\ref{Eq_cons_defor}),
  for some special families of undirected graphs. Average~consensus is achieved in all cases for~$s = 1$}\label{Table1}
\end{center}
\end{table*}
%
\begin{proposition}[$m$-cube $Q_m$]
For the $m$-cube (or~hypercube) graph $Q_m$ with $n = 2^m > 4$ 
vertices (see~Fig.~\ref{FIG:Graphs}(e)), we have that\footnote{$m$-cube or hypercube graphs should not
be confused with \emph{cubic graphs} which are 3-regular graphs. The
only hypercube which is a cubic graph is~$Q_3$.}: 
\begin{itemize}
\item For $|s| > 1$ or $|s| < \frac{1}{m-1}$, system~(\ref{Eq_cons_defor})
is asymptotically stable.
\item For $s \in \big(\!-1,\,-\frac{1}{m-1}\big)$ or $s \in \big(\frac{1}{m-1},\,1\big)$,
system (\ref{Eq_cons_defor}) is unstable.
\item For $s = \frac{1}{m-1}$, average consensus is achieved. The convergence rate
to $\frac{1}{n}\,\vet{x}_0^T\mathds{1}_{n}$ is slower for $s = \frac{1}{m-1}$ than for~$s = 1$.
\item For $s \in \{-1,\,-\frac{1}{m-1}\}$, system (\ref{Eq_cons_defor}) is marginally stable.
In this case, the states associated to $n/2$ vertices asymptotically converge to
$\frac{1}{\sqrt{n}}\,\vet{x}_0^T\vet{u}_1$, while the states associated to the other
$n/2$ vertices converge to $-\frac{1}{\sqrt{n}}\,\vet{x}_0^T\vet{u}_1$, 
where $\vet{u}_1$ is the unit-norm eigenvector associated to the zero eigenvalue
of $-\boldsymbol{\Delta}(-1)$ or $-\boldsymbol{\Delta}(-\frac{1}{m-1})$.
\end{itemize}\vspace{-0.2cm}
\emph{Proof}:
Since the eigenvalues of the adjacency matrix of the $m$-cube are the
numbers $m - 2\ell$ with multiplicity
$\left(\begin{smallmatrix}m\vspace{0.1cm}\\\ell \end{smallmatrix}\right)$, $\ell \in
\{0,\ldots,m\}$~\cite[Sect.~1.4.6]{BrouwerHa_book12}, then from
Lemma~\ref{Lemma1}, the eigenvalues of matrix $-\boldsymbol{\Delta}(s)$ 
are $-(m - 1)\,s^2 + (m - 2\ell)\,s - 1$ with multiplicity
$\left(\begin{smallmatrix}m\vspace{0.1cm}\\\ell \end{smallmatrix}\right)$. 
The stability of system~(\ref{Eq_cons_defor}) is only
determined by the eigenvalues, 
$$
\begin{array}{l}
\lambda_{1}(s) \,=\, -(m-1)\,s^2 \,+\, m\,s \,-\, 1,\vspace{0.14cm}\\
\lambda_{n}(s) \,=\, -(m-1)\,s^2 \,-\, m\,s \,-\, 1, \vspace{-0.1cm}
\end{array}
$$
since the other $n-2$ eigenvalues are negative for all $s$.
We have that $\lambda_{1}(s) < 0$ for $s > 1$ or $s < \frac{1}{m-1}$. Instead,
$\lambda_{n}(s) < 0$ for $s > -\frac{1}{m-1}$ or $s < -1$. Hence,  
for $|s| > 1$ or $|s| < \frac{1}{m-1}$, system~(\ref{Eq_cons_defor})
is asymptotically stable and for~$s \in \{-1,\,\pm \frac{1}{m-1}\}$,
marginal stability is achieved. Note in particular, that 
$\boldsymbol{\Delta}\big(\!-\!\frac{1}{m-1}\big) = \frac{1}{m-1}\,\boldsymbol{\Delta}(-1)$
and that for a suitable labeling of the vertices of the graph, we have 
$$
\!\lim\limits_{t \,\rightarrow \, \infty}
\,\,\exp(-\boldsymbol{\Delta}(-1)\,t)\,\vet{x}_0 \,=\, \frac{1}{n}\left[-\mathds{k},\,\mathds{k},\ldots,\,-\mathds{k},\,\mathds{k}\right]\vet{x}_0.
$$
Finally, note that $\boldsymbol{\Delta}\big(\frac{1}{m-1}\big) 
= \frac{1}{m-1}\,\boldsymbol{\Delta}(1) = \frac{1}{m-1}\,\vet{L}$ 
which implies that $\lambda_2\big(\boldsymbol{\Delta}\big(\frac{1}{m-1}\big)\big) 
\,=\, \frac{1}{m-1}\,\,\lambda_2\left(\boldsymbol{\Delta}(1)\right)$.
If we now recall the role played by the algebraic connectivity of the graph in the
consensus protocol, we have that
%
the convergence rate to $\frac{1}{n}\,\vet{x}_0^T\mathds{1}_{n}$
is slower for $s = \frac{1}{m-1}$ than for~$s = 1$.~\hfill$\blacksquare$ 
\end{proposition}
\vspace{-0.07cm}

\begin{proposition}[Petersen graph]
For the Petersen graph (see Fig.~\ref{FIG:Graphs}(f)), we have that:
\begin{itemize}
\item For $s > 1$ or $s < 1/2$, system (\ref{Eq_cons_defor}) is asymptotically stable.
\item For $s \in (1/2,\,1)$, system (\ref{Eq_cons_defor}) is unstable.
\item For $s = 1/2$, average consensus is achieved. The convergence rate
to $\frac{1}{10}\,\vet{x}_0^T\mathds{1}_{10}$ is slower for $s = 1/2$ than for~$s = 1$.\vspace{-0.12cm}
\end{itemize}
\emph{Proof}:
Since the eigenvalues of the adjacency matrix of the Petersen graph
are $3$, $1$ and $-2$ with multiplicity 1, 5 and 4, respectively~\cite[Sect.~1.4.5]{BrouwerHa_book12}, 
then from Lemma~\ref{Lemma1}, the eigenvalues of~$-\boldsymbol{\Delta}(s)$ are:
$$
\begin{array}{lll}
\displaystyle \lambda_{1}(s) &=& -2\,s^2 + 3\,s - 1,\vspace{0.08cm}\\
\displaystyle \lambda_{2}(s) &=& \ldots\, =\, \lambda_{6}(s) = -2\,s^2 + s - 1,\vspace{0.08cm}\\
\displaystyle \lambda_{7}(s) &=& \ldots\, =\, \lambda_{10}(s) = -2\,s^2 - 2\,s - 1.
\end{array}
$$
We have that $\lambda_{2}(s) < 0$ and $\lambda_{7}(s) < 0$, $\forall\, s \in \rr$. Moreover,
$\lambda_{1}(s) < 0$ for $s > 1$ or $s < 1/2$, and the unit-norm eigenvector associated
to $\lambda_{1}(1/2)$ is $\frac{1}{\sqrt{10}}\,\mathds{1}_{10}$.
\mbox{Finally}, $\boldsymbol{\Delta}(1/2)  =
\frac{1}{2}\,\boldsymbol{\Delta}(1)$, 
hence the convergence rate to $\frac{1}{10}\,\vet{x}_0^T\mathds{1}_{10}$
is slower for $s = 1/2$ than for~$s =~1$.
\hfill$\blacksquare$
\end{proposition}

\begin{proposition}[Complete graph $K_n$]
For the complete graph $K_n$ with $n > 2$ vertices (see Fig.~\ref{FIG:Graphs}(g)), we have~that:
\begin{itemize}
\item For $s > 1$ or $s < \frac{1}{n-2}$\,,\, system (\ref{Eq_cons_defor}) is asymptotically stable.
\item For $s \in \big(\frac{1}{n-2},\,1\big)$, system (\ref{Eq_cons_defor}) is unstable.
\item For $s = \frac{1}{n-2}$, average consensus is achieved.
The~convergence rate to $\frac{1}{n}\,\vet{x}_0^T\mathds{1}$
is slower for $s = \frac{1}{n-2}$ than for~$s = 1$.
\end{itemize}
\emph{Proof}:
Since the eigenvalues of the adjacency matrix of the complete graph
are $n-1$ and $-1$ with multiplicity $1$ and $n-1$, respectively~\cite[Sect.~1.4.1]{BrouwerHa_book12}, 
then from Lemma~\ref{Lemma1} the eigenvalues of
$-\boldsymbol{\Delta}(s)$ are: 
$$
\begin{array}{lll}
\displaystyle \lambda_{1}(s) &=& -(n-2)\,s^2 + (n-1)\,s - 1,\vspace{0.12cm}\\
\displaystyle \lambda_{2}(s) &=& \ldots\, = \; \lambda_{n}(s) \,=\, -(n-2)\,s^2 - s - 1.
\end{array}
$$
We have that $\lambda_{2}(s) < 0$, $\forall\, s \in \rr$. Moreover,
$\lambda_{1}(s) < 0$ for $s > 1$ or $s < \frac{1}{n-2}$, and the unit-norm eigenvector associated
to $\lambda_{1}\big(\frac{1}{n-2}\big)$ is $\frac{1}{\sqrt{n}}\,\mathds{1}$.
Finally, note that $\boldsymbol{\Delta}\big(\frac{1}{n-2}\big) 
= \frac{1}{n-2}\,\boldsymbol{\Delta}(1)$,
hence, the convergence rate to $\frac{1}{n}\,\vet{x}_0^T\mathds{1}$ is slower for
$s = \frac{1}{n-2}$ than for $s = 1$.
\hfill$\blacksquare$
\end{proposition}

\begin{proposition}[Complete bipartite graph $K_{m,n}$]\label{Prop_Com_bip}
For the complete bipartite graph $K_{m,n} = (V_1 \cup V_2,E)$,
where \mbox{$|V_1| = m$}, $|V_2| = n$ with $m,n \geq 2$ 
(see Fig.~\ref{FIG:Graphs}(h)), we have~that:
\begin{itemize}
\item For $|s| > 1$ or $|s| < \frac{1}{\sqrt{(m-1)(n-1)}}$, system (\ref{Eq_cons_defor}) is asymptotically stable.
\item For $s \in \big(-1, -\frac{1}{\sqrt{(m-1)(n-1)}}\big)$ or $s
  \in \big(\frac{1}{\sqrt{(m-1)(n-1)}}, 1\big)$, system (\ref{Eq_cons_defor}) is unstable.
\item For $s \in \big\{\!-1,\,\,\pm \frac{1}{\sqrt{(m-1)(n-1)}}\big\}$,
system (\ref{Eq_cons_defor}) is marginally stable.
In particular, given the initial condition \mbox{$\vet{x}_0 \in \rr^{m+n}$}:
\begin{itemize}
\item If $m \neq n$:\, for $s = -1$, the states associated to the vertices in $V_1$ asymptotically converge to
$\frac{1}{m+n}\,\vet{x}_0^T[\mathds{1}^T_m, -\!\mathds{1}^T_n]^T$ and
the states associated to the vertices in $V_2$ converge to $-\frac{1}{m+n}\,\vet{x}_0^T[\mathds{1}^T_m, -\!\mathds{1}^T_n]^T$.\\
For $s = \pm \frac{1}{\sqrt{(m-1)(n-1)}}$, the states associated to the vertices
in $V_1$ and $V_2$ asymptotically converge to one of the two different
values taken by the components of the vector $(\vet{u}_1\,\vet{u}^T_1)\,\vet{x}_0$, where 
$\vet{u}_1$ is the unit-norm eigenvector associated to the zero
eigenvalue of $-\boldsymbol{\Delta}\big(\pm\frac{1}{\sqrt{(m-1)(n-1)}}\big)$.
\item If $m = n$:\, for $s = \frac{1}{n-1}$ average consensus is achieved, and the convergence rate to
$\frac{1}{2n}\,\vet{x}_0^T\mathds{1}_{2n}$ is slower for $s = \frac{1}{n-1}$ than for \mbox{$s = 1$}.
For \mbox{$s \in \big\{\!\!-1, -\frac{1}{n-1}\big\}$} the states
associated to the vertices in $V_1$ asymptotically converge to 
$\frac{1}{2n}\,\vet{x}_0^T[\mathds{1}^T_n\!,-\mathds{1}^T_n]^T$ and the states associated to the vertices in $V_2$
converge to $-\frac{1}{2n}\,\vet{x}_0^T[\mathds{1}^T_n\!,-\mathds{1}^T_n]^T$.
\end{itemize}
\end{itemize}
\emph{Proof}:
In this case, the eigenvalues of $-\boldsymbol{\Delta}(s)$ are: 
$$
\begin{array}{l}
\displaystyle \lambda_{1}(s) = -\frac{n+m-2}{2}\,s^2 + \frac{\sqrt{(n-m)^2 s^2 +\, 4mn}}{2}\,s \,-\, 1,\vspace{0.42cm}\\
\displaystyle \lambda_{2}(s) = -\frac{n+m-2}{2}\,s^2 - \frac{\sqrt{(n-m)^2 s^2 +\, 4mn}}{2}\,s \,-\, 1,
\end{array}
$$
$$
\begin{array}{l}
\displaystyle \hspace{-1.2cm}\lambda_{3}(s) = \ldots = \lambda_{n+1}(s) = -((m-1)\,s^2 + 1),\vspace{0.18cm}\\
\displaystyle \hspace{-1.2cm}\lambda_{n+2}(s) = \ldots = \lambda_{n+m}(s) = -((n-1)\,s^2 + 1).
\end{array}
$$
Since $\lambda_{3}(s)$, $\ldots$, $\lambda_{n+m}(s)$ are negative for
all $s$, the stability of system~(\ref{Eq_cons_defor}) is only
determined by the eigenvalues $\lambda_{1}(s)$ and $\lambda_{2}(s)$.
A systematic study of the roots of $\lambda_{1}(s)$ and
$\lambda_{2}(s)$ immediately leads to the first two items of the
statement. For the marginal-stability case, note that for $s = -1$:\vspace{-0.2cm}
$$
\!\lim_{t \,\rightarrow \, \infty}\exp\big(-\boldsymbol{\Delta}(-1)\,t\big)\,\vet{x}_0 =
\frac{1}{m+n} \left[
\begin{matrix}
\mathds{J}_{m \times m} & -\mathds{J}_{m \times n}\\ 
-\mathds{J}_{n \times m} & \mathds{J}_{n \times n}
\end{matrix}
\right]\!\vet{x}_0.
$$
Moreover, observe that if $m=n$, $\boldsymbol{\Delta}\big(\!-\!\frac{1}{n-1}\big) 
= \frac{1}{n-1}\,\boldsymbol{\Delta}(-1)$, and analogously,
$\boldsymbol{\Delta}\big(\frac{1}{n-1}\big) 
= \frac{1}{n-1}\,\boldsymbol{\Delta}(1)$, 
hence the rate of convergence to $\frac{1}{2n}\,\vet{x}_0^T\mathds{1}_{2n}$ is slower
for $s = \frac{1}{n-1}$ than for~$s = 1$.~\hfill$\blacksquare$
\end{proposition}
From Prop.~\ref{Prop_Com_bip}, we can deduce the following 
result (recall~that the star graph is a complete bipartite graph with $m=1$):

\begin{proposition}[Star graph $K_{1,n}$]
For the star graph $K_{1,n}$ with $n \geq 3$, where vertex 1 is the
center of the star (see~Fig.~\ref{FIG:Graphs}(i)), we have that:
\begin{itemize}
\item For $|s| < 1$, system (\ref{Eq_cons_defor}) is asymptotically stable.
\item For $|s| > 1$, system (\ref{Eq_cons_defor}) is unstable.
\item For $s = -1$, system (\ref{Eq_cons_defor}) is marginally stable.
In this case, the state associated to vertex 1 asymptotically converges
to $\frac{1}{n+1}\,\vet{x}_0^T[1, -\mathds{1}^T_n]^T$
and the states associated to the other $n$ vertices 
converge to $-\frac{1}{n+1}\,\vet{x}_0^T[1, -\mathds{1}^T_n]^T$.~\hfill$\blacksquare$\vspace{0.05cm}
\end{itemize}
\end{proposition}

For the reader's convenience, all the results found in this section are summarized in Table~\ref{Table1}.

\begin{remark}\label{Remark_bipart}
Note that the path, cycle (with $n$ even), full $m$-ary tree, $m$-cube, complete
bipartite and star graphs are all \emph{bipartite} graphs. Then, in
view of Property~\ref{Prop_Q}.6, for $s = -1$ the state of system~(\ref{Eq_cons_defor})
asymptotically converges to $(\vet{u}_n\vet{u}_n^T)\,\vet{x}_0$ with these graphs, where $\vet{u}_n$
is the unit-norm eigenvector associated to the zero eigenvalue of the signless
Laplacian~$\vet{Q}$. According to~\cite[Def.~1]{Altafini_TAC13}, 
we can say that system~(\ref{Eq_cons_defor}) admits a \emph{bipartite
consensus solution} in these cases.~\hfill$\diamond$ 
\end{remark}


\subsection{Stability conditions for graphs of arbitrary topology}\label{Sect:Gen}

In order to extend the analysis of the previous section to arbitrary undirected graphs, 
we briefly review here the spectral theory of \emph{quadratic eigenvalues problems}
(QEPs)~\cite[Sect.~3]{TisseurMe_SIAM01},
that constitute an important class of nonlinear eigenvalue problems. Let
$$
\vet{P}(\lambda) \,=\, \vet{B}_2\,\lambda^2 + \vet{B}_1\,\lambda + \vet{B}_0,
$$
be an $n \times n$ \emph{matrix polynomial} of degree 2,
where $\vet{B}_2$, $\vet{B}_1$ and $\vet{B}_0$ are $n \times n$ complex matrices~\cite{GohbergLaRo_Book09}.
In other words, the components of the matrix $\vet{P}(\lambda)$ are
quadratic polynomials in the variable $\lambda$.

\begin{definition}[Spectrum of $\vet{P}(\lambda)$]
The \emph{spectrum} of $\vet{P}(\lambda)$, denoted by $\Sigma(\vet{P})$, is defined as,
$$
\Sigma(\vet{P}) = \{\,\lambda \in \mathbb{C}\,:\, \det(\vet{P}(\lambda)) = 0\},
$$
i.e., it is the set of eigenvalues of $\vet{P}(\lambda)$~\cite{TisseurMe_SIAM01}.\hfill$\diamond$
\end{definition}

\begin{definition}[Regular $\vet{P}(\lambda)$]\label{Def_regular}
The matrix $\vet{P}(\lambda)$ is called \emph{regular} when $\det(\vet{P}(\lambda))$ is not identically zero
for all values of $\lambda$, and \emph{nonregular} otherwise~\cite{TisseurMe_SIAM01}.\hfill$\diamond$
\end{definition}

Note that $\det(\vet{P}(\lambda)) = \det(\vet{B}_2)\lambda^{2n} +$
lower-order terms, 
so when $\vet{B}_2$ is nonsingular, $\vet{P}(\lambda)$ is regular and has
$2n$ \emph{finite} eigenvalues. When $\vet{B}_2$ is singular, the degree of
$\det(\vet{P}(\lambda))$ is $r < 2n$ and $\vet{P}(\lambda)$ has $r$ finite eigenvalues and
$2n - r$ infinite eigenvalues\footnote{The infinite eigenvalues correspond to the zero eigenvalues
of the reverse polynomial $\lambda^2\,\vet{P}(\lambda^{-1}) =
\vet{B}_0\,\lambda^2 + \vet{B}_1\,\lambda + \vet{B}_2$.}.
The \emph{algebraic multiplicity} of an eigenvalue $\lambda_0$ is the order of the corresponding
zero in $\det(\vet{P}(\lambda))$, while the \emph{geometric multiplicity} of
$\lambda_0$ is the dimension of $\ker(\vet{P}(\lambda_0))$.

\begin{problem}[Quadratic eigenvalue problem, QEP]
The QEP consists of finding scalars $\lambda$ and nonzero vectors $\vet{z}$, $\vet{y}$,
satisfying~\cite{TisseurMe_SIAM01},
$$
\vet{P}(\lambda)\,\vet{z} = \vet{0},\quad \,\vet{y}^*\,\vet{P}(\lambda) = \vet{0},
$$
where $\vet{z}$, $\vet{y} \in \mathbb{C}^n$ are respectively the right and left eigenvector corresponding
to the eigenvalue $\lambda \in \mathbb{C}$, and $\vet{y}^*$ is the conjugate transpose of $\vet{y}$.~\hfill$\diamond$
\end{problem}

A QEP has $2n$ eigenvalues (finite or infinite) with up to $2n$ right and $2n$ left eigenvectors.
Note that a regular $\vet{P}(\lambda)$ may possess two distinct eigenvalues having the same eigenvector.
If a regular $\vet{P}(\lambda)$ has $2n$ distinct eigenvalues, then there exists a set
of $n$ linearly independent eigenvectors.

\begin{property}[Spectral properties of\, $\vet{P}(\lambda)$]\label{Prop_Q_lambda}
If~matrices $\vet{B}_2$, $\vet{B}_1$, $\vet{B}_0$ are \emph{real symmetric}, the eigenvalues of $\vet{P}(\lambda)$
are either real or occur in complex-conjugate pairs, and the sets of
left and right eigenvectors coincide~\cite{TisseurMe_SIAM01}.~\hfill$\diamond$
\end{property}

By leveraging the previous facts (note that according to Def.~\ref{Def_regular}, 
$\boldsymbol{\Delta}(\lambda)$ is a regular matrix since we always
have $\det(\boldsymbol{\Delta}(\lambda)) \neq 0$ for $\lambda = 0$),
we deduce the following property of
the deformed consensus protocol~(\ref{Eq_cons_defor}):

\begin{proposition}\label{Prop_Main}
The \emph{finite real eigenvalues} $\lambda$ of the~QEP,
\begin{equation}\label{Eq:QEP:me}
((\vet{I}_n - \vet{D})\,\lambda^2 + \vet{A}\,\lambda - \vet{I}_n)\,\vet{z} \,=\, \vet{0},
\end{equation}
are the values of $s$ for which system~(\ref{Eq_cons_defor}) is marginally stable. Moreover, if
$\lambda$ is one of these eigenvalues with geometric
multiplicity one, and $\overline{\vet{z}} = \vet{z}/\|\vet{z}\|$ is the associated
unit-norm eigenvector, we have that:
$$
\hspace{2.4cm}\lim_{t \rightarrow \infty}\,\vet{x}(t) \;=\; (\overline{\vet{z}}\,\,\overline{\vet{z}}^T)\,\vet{x}_0.
\hspace{2.2cm}\hfill\blacksquare
$$
\end{proposition}
%
%

\begin{figure}[t!]
       \begin{center}
        \psfrag{a}{$s$}
        \psfrag{b}{$q(s)$}
        \psfrag{c}{\scriptsize{asymptotic stability}}
        \psfrag{d}{\tiny{unstab.}}
        \psfrag{e}{\scriptsize{unstab.}}
        \psfrag{f}{\scriptsize{marginal stability}}
       \includegraphics[width=.85\columnwidth]{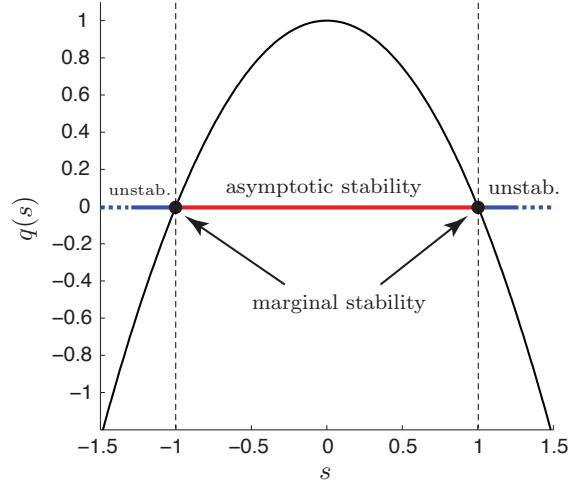}
       \vspace{-0.15cm} 
       \caption{\emph{Illustration of Prop.~\ref{Prop_stab_gen}}: For the path graph $P_6$, $q(s) = 1-s^2$.}\label{Fig_det_Delta}
       \end{center}
\end{figure}

\begin{remark}[Computation of the eigenvalues] 
The eigenvalues of the QEP~(\ref{Eq:QEP:me}) can be easily computed by converting
it to a standard \emph{generalized eigenvalue problem}\footnote{This construction is called ``linearization''
in the literature and it is not unique in general (see~\cite[Sect.~4.5]{Demmel_book97}).} of size $2n$, by defining the new
vector $\vet{w} = \lambda\,\vet{z}$. In~terms of $\vet{z}$ and
$\vet{w}$, problem~(\ref{Eq:QEP:me}) becomes:
$$
\qquad\;\left[\,
  \begin{matrix}
    \vet{0} \;&\; \vet{I}_n\\
    \vet{I}_n \;&\; -\vet{A} \\
  \end{matrix}
\,\right]\!\left[\,
         \begin{matrix}
           \vet{z}\\
           \vet{w}
         \end{matrix}
       \,\right] \,=\, \lambda\!\left[\,
  \begin{matrix}
    \vet{I}_n \;&\; \vet{0}\\
    \vet{0} \;&\; \vet{I}_n - \vet{D} \\
  \end{matrix}
\,\right]\!\left[\,
         \begin{matrix}
           \vet{z}\\
           \vet{w}
         \end{matrix}
 \,\right]\!.
$$
Matlab's function ``\texttt{polyeig}'' uses a linearization procedure
similar to the one described above, to numerically solve 
generic polynomial eigenvalue problems. Note that if matrix $\vet{I}_n -
\vet{D}$ is singular, it is more convenient, from a numerical
viewpoint, to use the second instead of the first companion form
considered above, for the linearization (cf.~\cite[Sect.~3.4]{TisseurMe_SIAM01}).~\hfill$\diamond$
\end{remark}
The following proposition elucidates the connection existing \mbox{between} the topology of the
communication graph~$\mathcal{G}$, and the properties of the QEP~(\ref{Eq:QEP:me}).
\begin{proposition}
If the graph $\mathcal{G}$ has a vertex with degree equal to one (i.e. only one edge
is incident to that vertex), 
$\vet{I}_n - \vet{D}$ is singular and the QEP~(\ref{Eq:QEP:me}) admits at least two infinite eigenvalues.~\hfill$\blacksquare$
\end{proposition}
For example, with the path graph $\text{rank}(\vet{I}_n - \vet{D}) = n-2$,
with the star graph $\text{rank}(\vet{I}_n - \vet{D}) = 1$ and with the full
$m$-ary tree graph $\text{rank}(\vet{I}_n - \vet{D}) = n - m^{\delta}$:
in the first two cases $\det((\vet{I}_n - \vet{D})\lambda^2 + \vet{A}\,\lambda - \vet{I}_n) = (-1)^{n-1}(\lambda^2 - 1)$,
while in the last $\det((\vet{I}_n - \vet{D})\lambda^2 + \vet{A}\,\lambda - \vet{I}_n) = \lambda^2 - 1$.
Hence, in all cases, the QEP (\ref{Eq:QEP:me})~admits $2n - 2$ infinite eigenvalues.\\
The next proposition, the last result of this section, shows how to determine the \mbox{$s$-stability} interval of
the deformed consensus protocol, for graphs with arbitrary
\mbox{topology} (see Fig.~\ref{Fig_det_Delta}).
%
\begin{figure*}[t!]
       \begin{center}
       \begin{tabular}{cccc}
       \!\subfigure[]{\includegraphics[width=.428\columnwidth]{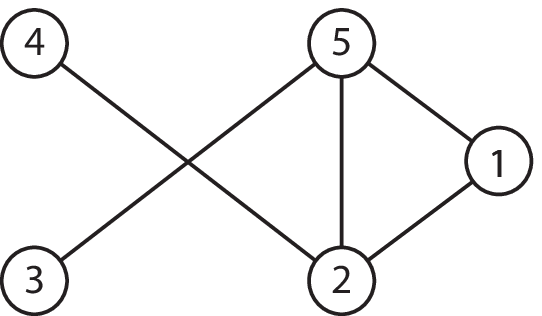}}
       \hspace{0.17cm}&\hspace{0.17cm}
       \subfigure[]{\includegraphics[width=.428\columnwidth]{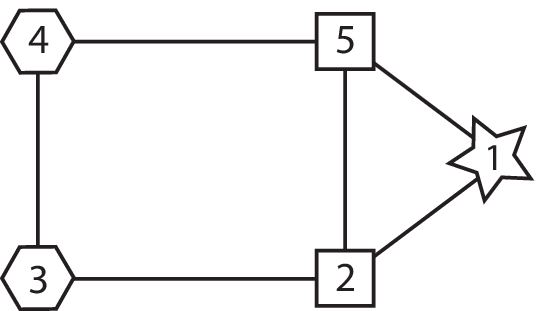}}
       \hspace{0.17cm}&\hspace{0.17cm}
       \subfigure[]{\includegraphics[width=.428\columnwidth]{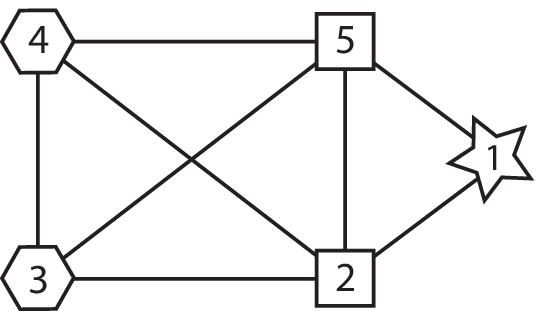}}
       \hspace{0.17cm}&\hspace{0.17cm}
       \subfigure[]{\includegraphics[width=.428\columnwidth]{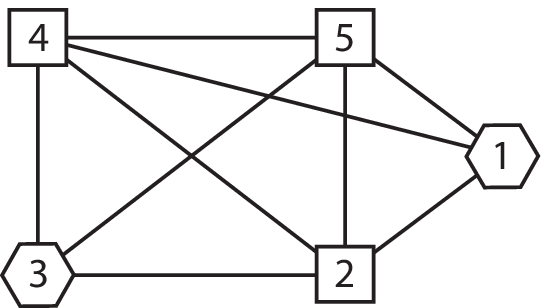}}
       \end{tabular}
           \caption{\emph{Example 1}: In (b)-(d), different shapes are used to identify distinct
           groups of vertices: the states associated to the vertices in these groups asymptotically
           converge to the same value when system~(\ref{Eq_cons_defor}) is marginally~stable.}
           \label{FIG:Example}
       \end{center}
\end{figure*}
\begin{proposition}[Stability interval for arbitrary $\mathcal{G}$]\label{Prop_stab_gen} 
Let $q(s) \,\triangleq\, \det((\vet{I}_n - \vet{D})\,s^2 + \vet{A}\,s - \vet{I}_n)$,
then:
\begin{itemize}
\item If $n$ is \emph{even}, system~(\ref{Eq_cons_defor}) is
  asymptotically stable for all~$s$ such that $q(s) > 0$, and unstable for all $s$ such that $q(s) < 0$.
\item If $n$ is \emph{odd}, system~(\ref{Eq_cons_defor}) is
  asymptotically stable for all~$s$ 
such that $q(s) < 0$, and unstable for all $s$ such that $q(s) > 0$.
\end{itemize}
\vspace{-0.1cm}
\emph{Proof}:
From Property~\ref{Prop_Q_lambda} and Property~\ref{Sp_propL}.1, 
it follows that:\vspace{-0.07cm} 
$$
q(s) \,=\, 
(s - 1)\prod_{k=1}^{\ell} (s - \zeta_k)(s - \zeta_k^*)\prod_{j\,=\,1}^{r -\, 2\ell -1} (s - \eta_j),
$$
where $r \leq 2n$ is the degree of $q(s)$, $(\zeta_k,\, \zeta_k^*)$
are the $\ell$~pairs of complex-conjugate roots of $q(s)$ and $\eta_j$ the $r - 2\ell - 1$ non-unitary real
roots of $q(s)$ (counted with their multiplicity). Since $\prod_{k=1}^{\ell} (s - \zeta_k)(s
- \zeta_k^*) = \prod_{k=1}^{\ell} s^2 - 2 \Re[\zeta_k]\,s + |\zeta_k|^2 > 0$ for all $s$,
where $\Re[\,\cdot\,]$ denotes the real part of a complex number,
the statement follows from Prop.~\ref{Prop_Main}.~\hfill$\blacksquare$
%
\end{proposition}
The following example illustrates the rich variety of behaviors exhibited by the deformed consensus protocol
on four ``generic'' (nonbipartite) graphs with five vertices.
\begin{example}\label{Example}
Consider the four graphs reported in~Fig.~\ref{FIG:Example}. By leveraging Prop.~\ref{Prop_Main}
and Prop.~\ref{Prop_stab_gen}, we have that:
\begin{itemize}
\item With the graph in Fig.~\ref{FIG:Example}(a), system~(\ref{Eq_cons_defor})
is asymptotically stable $\forall\, s \in \rr \setminus \{1\}$.
\item With the graph in Fig.~\ref{FIG:Example}(b), system~(\ref{Eq_cons_defor})
is asymptotically stable for $s < 0.7022$ or $s > 1$. For $s = 0.7022$, the
system is marginally stable and three groups of vertices can be identified:
$\{1\}$, $\{2,5\}$, $\{3,4\}$ (different shapes are used in Fig.~\ref{FIG:Example}(b)
to indicate these groups).
\item With the graph in Fig.~\ref{FIG:Example}(c), system~(\ref{Eq_cons_defor})
is asymptotically stable for $s < 0.4396$ or $s > 1$. For $s = 0.4396$,
the system is marginally stable and three groups of vertices can be identified: $\{1\}$, $\{2,5\}$, $\{3,4\}$.
\item With the graph in Fig.~\ref{FIG:Example}(d), system~(\ref{Eq_cons_defor})
is asymptotically stable for $s < 0.3804$ or $s > 1$. For $s = 0.3804$,
the system is marginally stable and two groups of vertices can be identified: $\{1,3\}$, $\{2,4,5\}$.
\end{itemize}
From Figs.~\ref{FIG:Example}(b)-(d), we notice that vertices in the same
group tend to have the same edge degree,
and that an increase in the algebraic connectivity of the graph leads
to a shrinkage of the $s$-stability interval of the deformed consensus
protocol. In future works, we will delve into 
the peculiar grouping behavior exhibited by the vertices of the four
graphs considered in this example.~\hfill$\diamond$
\end{example}
%
%

\section{Extensions}\label{Sect:ext}

In this section, two extensions to the theory presented in
Sect.~\ref{SEC:Prob} are discussed: we first briefly consider
a \mbox{discrete-time version} of the deformed consensus 
protocol~(\ref{Eq_cons_defor}), and then deal with directed 
communication graphs.

\subsection{Discrete-time deformed consensus protocol}\label{Sect:ext_DT}

Following~\cite[Sect.~IIC]{OlfatiFaMu_IEEE07}, we can introduce the following discrete-time
version of protocol~(\ref{Eq_cons_defor}),
\begin{equation}\label{Eq:Def_cons_discr}
\vet{x}(k+1) \,=\, \mathds{P}(s)\,\vet{x}(k),\;\; k \in \{0,\,1,\,2,\ldots\},
\end{equation}
where
$$
\mathds{P}(s) \,=\, \vet{I}_n - \epsilon\,\boldsymbol{\Delta}(s),
$$
is the \emph{deformed Perron matrix} and $0 < \epsilon < 1/d_{\max}$ is the step-size,
where $d_{\max} = \max_{\,i}\, (\sum_{j \neq i} a_{ij})$ is the maximum
degree of $\mathcal{G}$.
It is easy to verify that the continuous-time and discrete-time deformed consensus protocols
share the same $s$-stability intervals, and thus the analysis of
Sect.~\ref{SEC:Prob}, \emph{mutatis mutandis}, is still valid
for system~(\ref{Eq:Def_cons_discr}).

\subsection{Deformed consensus protocol for directed graphs}\label{Sect:ext_Direct}

In this section we assume that the communication graph is directed 
and contains a rooted out-branching, and by mimicking Sect.~\ref{SEC:Prob} we will study the stability properties
of the following linear system,
\begin{equation}\label{Eq_cons_defor_dir}
\dot{\vet{x}}(t) = -\boldsymbol{\Delta}(\mathcal{D}(s))\,\vet{x}(t),
\end{equation}
in terms of the real parameter $s$, where the symbol ``$\mathcal{D}(s)$''
indicates that the deformed Laplacian is now relative to a directed communication topology.
Similarly to Sect.~\ref{SEC:Form_Prob}, we have here that:
$$
\boldsymbol{\Delta}(\mathcal{D}(1)) \,=\, \vet{L}(\mathcal{D}),\quad\; \boldsymbol{\Delta}(\mathcal{D}(-1)) \,=\, \vet{Q}(\mathcal{D}).
$$
It is well known~\cite{MesbahiEg_book10}, that if the digraph $\mathcal{D}$ contains a rooted out-branching, the
state trajectory of system,
\begin{equation}\label{Eq_cons_dir}
\dot{\vet{x}}(t) = -\vet{L}(\mathcal{D})\,\vet{x}(t),
\end{equation}
satisfies,
$$
\lim_{t \rightarrow \infty}\; \vet{x}(t) \,=\, (\overline{\vet{u}}_1 \overline{\vet{v}}^T_1)\,\vet{x}_0,
$$
where $\overline{\vet{u}}_1$ and $\overline{\vet{v}}_1$, are, respectively, the right and
left eigenvectors associated with the zero eigenvalue
of $\vet{L}(\mathcal{D})$, normalized such that $\overline{\vet{u}}^T_1\overline{\vet{v}}_1 = 1$. Moreover, we have
that~(\ref{Eq_cons_dir}) reaches \emph{average consensus} for every
initial state if and only if $\mathcal{D}$ is weakly connected and balanced.\\
In what follows, we will analyze the stability properties of system~(\ref{Eq_cons_defor_dir}) for
two special families of directed graphs, and briefly explore the case of digraphs of arbitrary topology with
the help of few significative examples.

\begin{figure}[t!]
       \begin{center}
       \begin{tabular}{cc}
       \!\!\!\psfrag{n}{\hspace{-0.045cm}\footnotesize{$n$}}
       \subfigure[]{\includegraphics[width=.51\columnwidth]{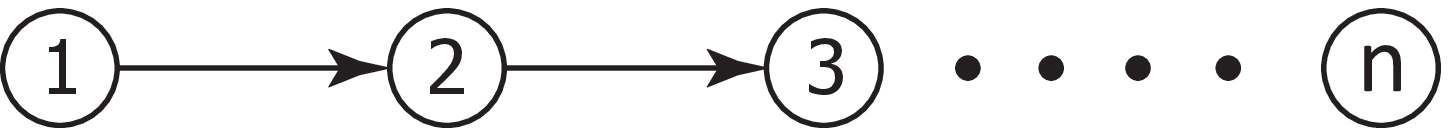}} &
       \psfrag{n}{\hspace{-0.045cm}$n$}
       \subfigure[]{\includegraphics[width=.44\columnwidth]{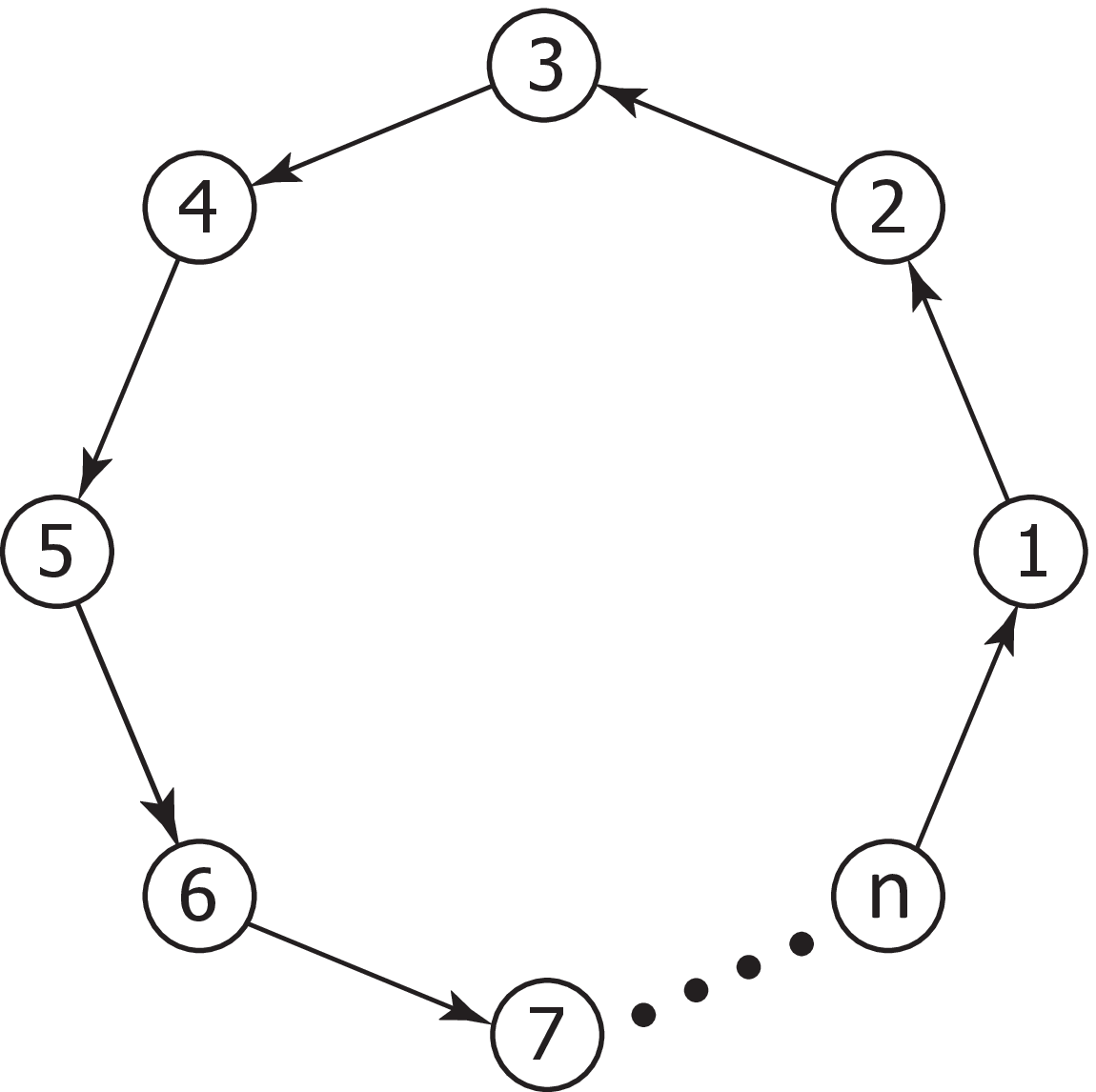}}
       \end{tabular}
       \vspace{-0.08cm}
       \caption{\emph{Two families of directed graphs}: (a) Directed path; (b) Directed cycle. Note that
       the former is weakly connected and not balanced, and the latter is strongly connected and balanced.}
       \label{FIG:Directed_graphs}
       \end{center}
\end{figure}

\begin{proposition}[Directed path]\label{Prop_DirPath}
For the directed path graph with $n \geq 2$ vertices (see Fig.~\ref{FIG:Directed_graphs}(a)), we have~that:
\begin{itemize}
\item For $|s| < 1$, system (\ref{Eq_cons_defor_dir}) is asymptotically stable.
\item For $|s| > 1$, system (\ref{Eq_cons_defor_dir}) is unstable.
\item For $s \in \{-1,\,+1\}$, system (\ref{Eq_cons_defor_dir}) is marginally stable.
\begin{itemize}
\item For $s = -1$, it is possible to identify two groups of $n/2$ vertices
(if $n$ is even), or one group of $\lfloor n/2 \rfloor$ vertices
and one of $\lfloor n/2 \rfloor + 1$ vertices (if $n$ is odd).
The states associated to the vertices in one group reach an agreement on $x_1(0)$,
and the states associated to the vertices in the other group
agree on $-x_1(0)$.
\item For $s = 1$, the consensus value is $x_1(0)$.
\end{itemize}
\end{itemize}
\emph{Proof}:
In this case, $-\boldsymbol{\Delta}(\mathcal{D}(s))$ is a lower-triangular \mbox{matrix},\vspace{-0.23cm}
$$
-\boldsymbol{\Delta}(\mathcal{D}(s)) \,= \left[\,
  \begin{matrix}
     s^2 \!-\! 1 & & & & &\vspace{-0.15cm}\\
     s &\, -1 & & & &\vspace{-0.1cm}\\
     & & &\ddots & &\vspace{-0.15cm}\\
     & & & s\;\; &\;\, -1 &\vspace{-0.15cm}\\
     & & & & s\;\; &\;\, -1
  \end{matrix}
\,\right],
$$
whose eigenvalues are,
$$
\lambda_{1}(s) = s^2-1,\quad \lambda_2(s) =\, \ldots \,= \lambda_{n}(s) = -1,
$$
from which the first two items of the statement \mbox{immediately}
follow. For the marginal-stability case, note~that,
$$
\lim_{t \,\rightarrow \, \infty}\; \exp(-\boldsymbol{\Delta}(\mathcal{D}(-1))\,t)\,\vet{x}_0 \,=\,
\big[\!-\mathds{k},\,\,\vet{0}_{n \times (n-1)}\big]\,\vet{x}_0,
$$
and
$$
\lim_{t \,\rightarrow \, \infty} \exp(-\boldsymbol{\Delta}(\mathcal{D}(1))\,t)\,\vet{x}_0 =
\big[\mathds{1},\,\vet{0}_{n \times (n-1)}\big]\,\vet{x}_0 = x_1(0)\,\mathds{1}.\vspace{-0.3cm}
$$
\hfill$\blacksquare$
\end{proposition}
Protocol~(\ref{Eq_cons_defor_dir}) exhibits a richer set of behaviors 
with directed cycle graphs, as detailed in the next proposition. 
\begin{table*}[t!]
\renewcommand{\arraystretch}{1.42}
\begin{center}
  \begin{tabular}{|l|c|c|} 
    \hline
    \textbf{Digraph name} & \textbf{Asymptotic stability for\,:} & \textbf{Marginal stability for\,:}\\
    \hline\hline
    Directed path, $n \geq 2$ & $|s| < 1$ & 
    $
    \begin{array}{c}
    s = -1\;\, \text{(2 groups of vertices)}\\
    s = 1 \;\, \text{(consensus)}
    \end{array}
    $\vspace{0.01cm}\\
    \hline
    Directed cycle $\emph{\textsf{D}}_n$, $n > 2$, $n$ even & $|s| <
    1$ &  
    $
    \begin{array}{c}
    s = -1\;\, \text{(2 groups of vertices)}\\
    s = 1 \;\, \text{(average consensus)}
    \end{array}
    $ \vspace{0.01cm}\\
    \hline
    Directed cycle $\emph{\textsf{D}}_n$, $n > 2$, $n$ odd & $s \in
    (\vartheta(n),\,1)$ & 
    $
    \begin{array}{c}
    s = \vartheta(n)\;\, \text{(stable oscillations)}\\ 
    s = 1\;\, \text{(average consensus)}
    \end{array}
    $ \vspace{0.01cm}\\
    \hline
  \end{tabular}
  \vspace{0.4cm}
  \caption{Summary of the stability properties of the deformed consensus protocol~(\ref{Eq_cons_defor_dir}),
  for two families of directed graphs.}\label{Table2}
\end{center}
\end{table*}
\begin{proposition}[Directed cycle $\textsf{D}_n$]\label{Prop_DirCyc}
For the directed cycle graph $\textsf{D}_n$ with $n > 2$ vertices 
(see Fig.~\ref{FIG:Directed_graphs}(b)), we have~that:
\begin{itemize}
\item For $s = 1$ and for all $n > 2$, average consensus is achieved.
\item If $n$ is even:
\begin{itemize}
\item For $|s| < 1$, system~(\ref{Eq_cons_defor_dir}) is asymptotically stable.
\item For $|s| > 1$, system~(\ref{Eq_cons_defor_dir}) is unstable.
\item For $s = -1$, system (\ref{Eq_cons_defor_dir}) is marginally stable.
In this case, the states associated to $n/2$ vertices asymptotically
converge to $\frac{1}{n}\,\vet{x}_0^T\,\mathds{k}$
and the states associated to the other $n/2$ vertices 
converge to $-\frac{1}{n}\,\vet{x}_0^T\,\mathds{k}$.
\end{itemize}
\item If $n$ is odd:
\begin{itemize}
\item System (\ref{Eq_cons_defor_dir}) is asymptotically stable for $s \in (\vartheta(n),\, 1)$, where,
$$
\vartheta(n) \,=\, \frac{1}{\displaystyle\cos\!\left(\frac{n(n-2) + 1}{n}\,\pi\right)}\,.
$$
\item For $s < \vartheta(n)$ or $s > 1$, system (\ref{Eq_cons_defor_dir}) is unstable.
\item For $s = \vartheta(n)$, system~(\ref{Eq_cons_defor_dir}) is marginally stable. At~steady-state,
we have that the $i$-th state of system~(\ref{Eq_cons_defor_dir}), $i
\in \{1,\ldots,n\}$, obeys, 
$$
x_i(t) \,=\, \textup{A} \sin\!\big(2\pi\,f(n)\,t \,+\, \phi_i(n) \,+\, \phi_{\circ}\big), 
$$
where $\textup{A}$ and $\phi_{\circ}$ are positive constants, the frequency
$$
f(n) \,=\, \frac{1}{2\pi}\,\tan\!\left(\frac{n(n-2) + 1}{n}\,\,\pi\!\right)\!,
$$
and for $i \in \{1,\ldots,n\}$, the phase
$$
\phi_i(n)\,=\, \frac{2\pi\,(i-1)}{n \tan\!\Big(\frac{n(n-2) + 1}{n}\,\pi\Big)}\,.
$$
\end{itemize}
\end{itemize}
\emph{Proof}:
Similarly to Prop.~\ref{Prop_Cycle}, $-\boldsymbol{\Delta}(\mathcal{D}(s))$
is a \emph{circulant matrix} in this case,
$$
-\boldsymbol{\Delta}(\mathcal{D}(s)) \,=\, \text{circ}[-1,\,0,\,\ldots,\,0,\,s],
$$
and using formula~(\ref{Eq:Eig_circ}) the eigenvalues of $-\boldsymbol{\Delta}(\mathcal{D}(s))$
can be computed in closed-form as,
\begin{equation}\label{Eq_eig_DC}
\lambda_i(s) \,=\, s\,\exp\!\left(\frac{2\pi(i-1)(n-1)j}{n}\right)\, -1,\;\; i \in \{1,\ldots,\,n\}, 
\end{equation}
where $j$ is the imaginary unit. The statement easily follows from a systematic
study of~(\ref{Eq_eig_DC}) in terms of parameter~$s$. In particular, for $s = 1$,
$\boldsymbol{\Delta}(\mathcal{D}(s))$ reduces to the standard 
Laplacian matrix and being $\textsf{D}_n$ strongly connected and balanced,
average consensus is achieved. \mbox{If $n$ is even} and $s = -1$, 
$$
\lim_{t \,\rightarrow \, \infty} \exp(-\boldsymbol{\Delta}(\mathcal{D}(-1))\,t)\,\vet{x}_0 =
\frac{1}{n}\left[-\mathds{k},\,\mathds{k},\,\ldots,\,-\mathds{k},\,\mathds{k}\right]\vet{x}_0.
$$
On the other hand, if $n$ is odd and $s = \vartheta(n)$, the following
pair of \emph{purely imaginary} eigenvalues appears in~(\ref{Eq_eig_DC}),
$$
\lambda_{\lceil n/2 \rceil,\, \lceil n/2 \rceil + 1} \,=\, \pm\, j\,\tan\!\left(\frac{n(n-2) + 1}{n}\,\pi\right),
$$
where $\lceil \cdot \rceil$ denotes the ceiling function, all the
other eigenvalues having negative real parts.
$\lambda_{\lceil n/2 \rceil}$, $\lambda_{\lceil n/2
  \rceil + 1}$ are responsible for the periodic solutions of period,
$$
\frac{1}{f(n)}\,=\, \frac{2\pi}{\tan\!\Big(\frac{n(n-2) + 1}{n}\,\pi\Big)},
$$
and phase $\phi_i(n)$ of system~(\ref{Eq_cons_defor_dir}), (cf.~\cite[p.~134]{Strogatz_book94}).
\hfill$\blacksquare$
\end{proposition}
\begin{figure*}[t!]
       \begin{center}
       \begin{tabular}{cccc}
       \!\!\subfigure[]{\includegraphics[width=.427\columnwidth]{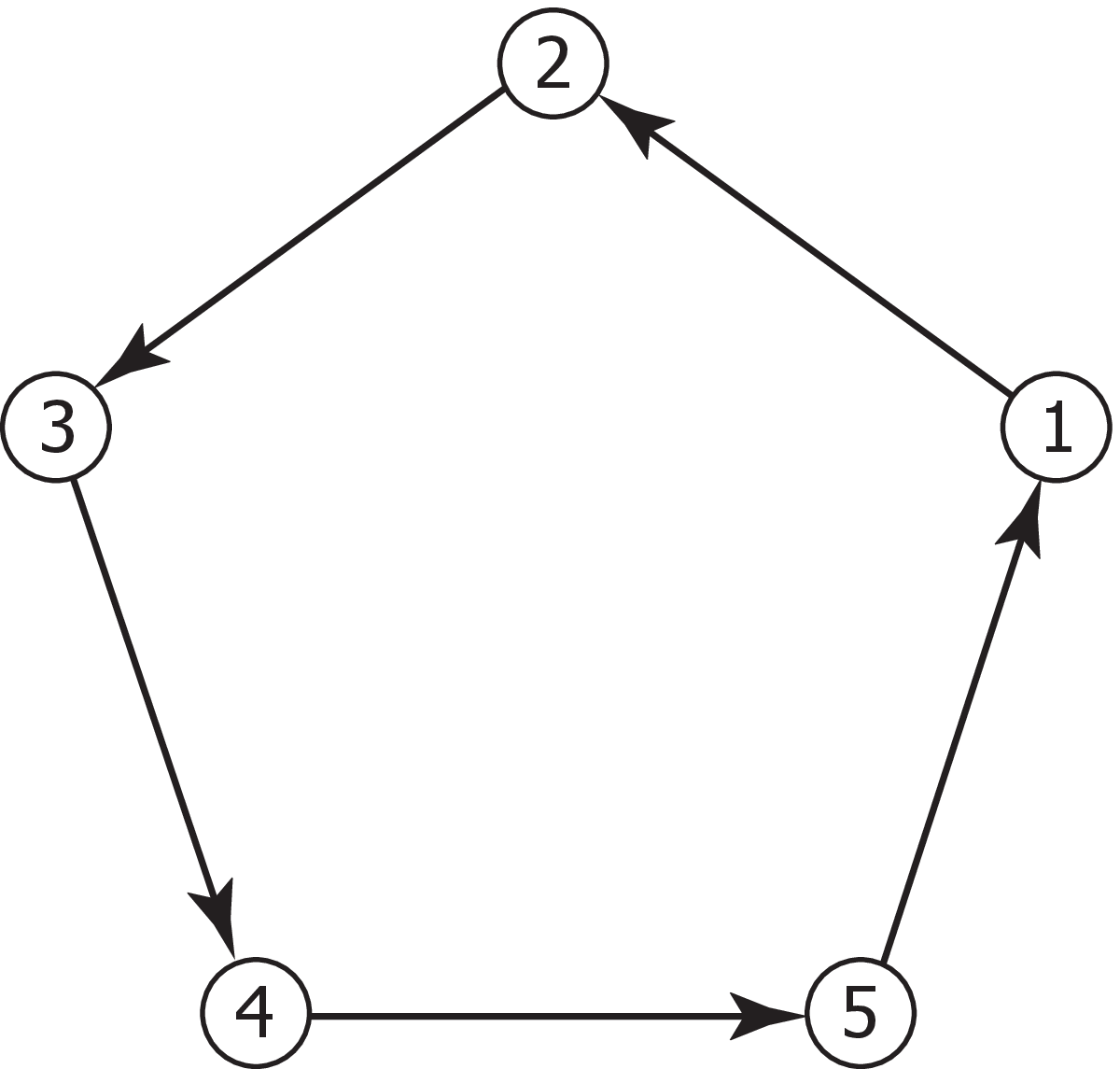}} \;\;&\;\;
       \subfigure[]{\includegraphics[width=.427\columnwidth]{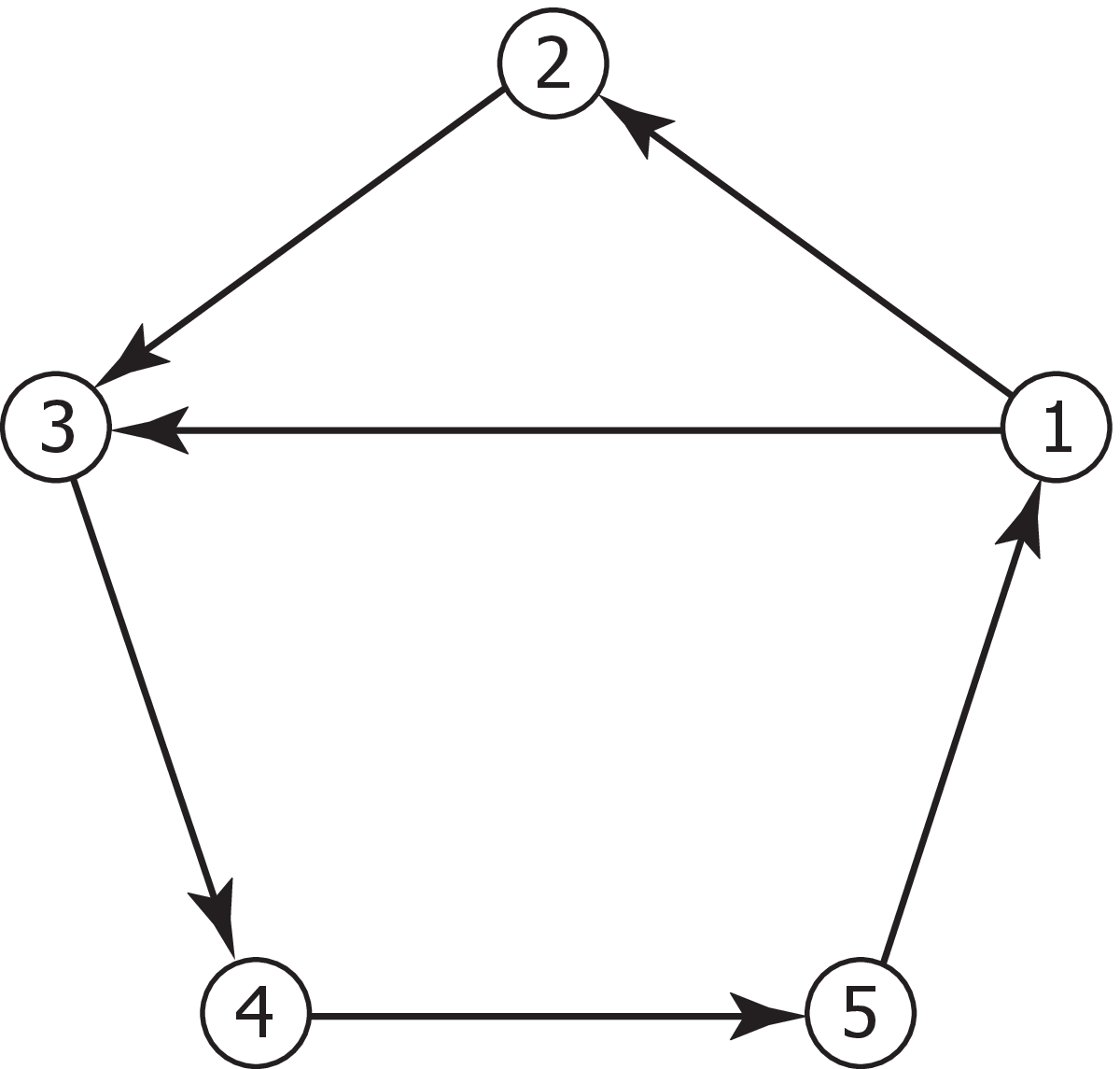}} \;\;&\;\;
       \subfigure[]{\includegraphics[width=.427\columnwidth]{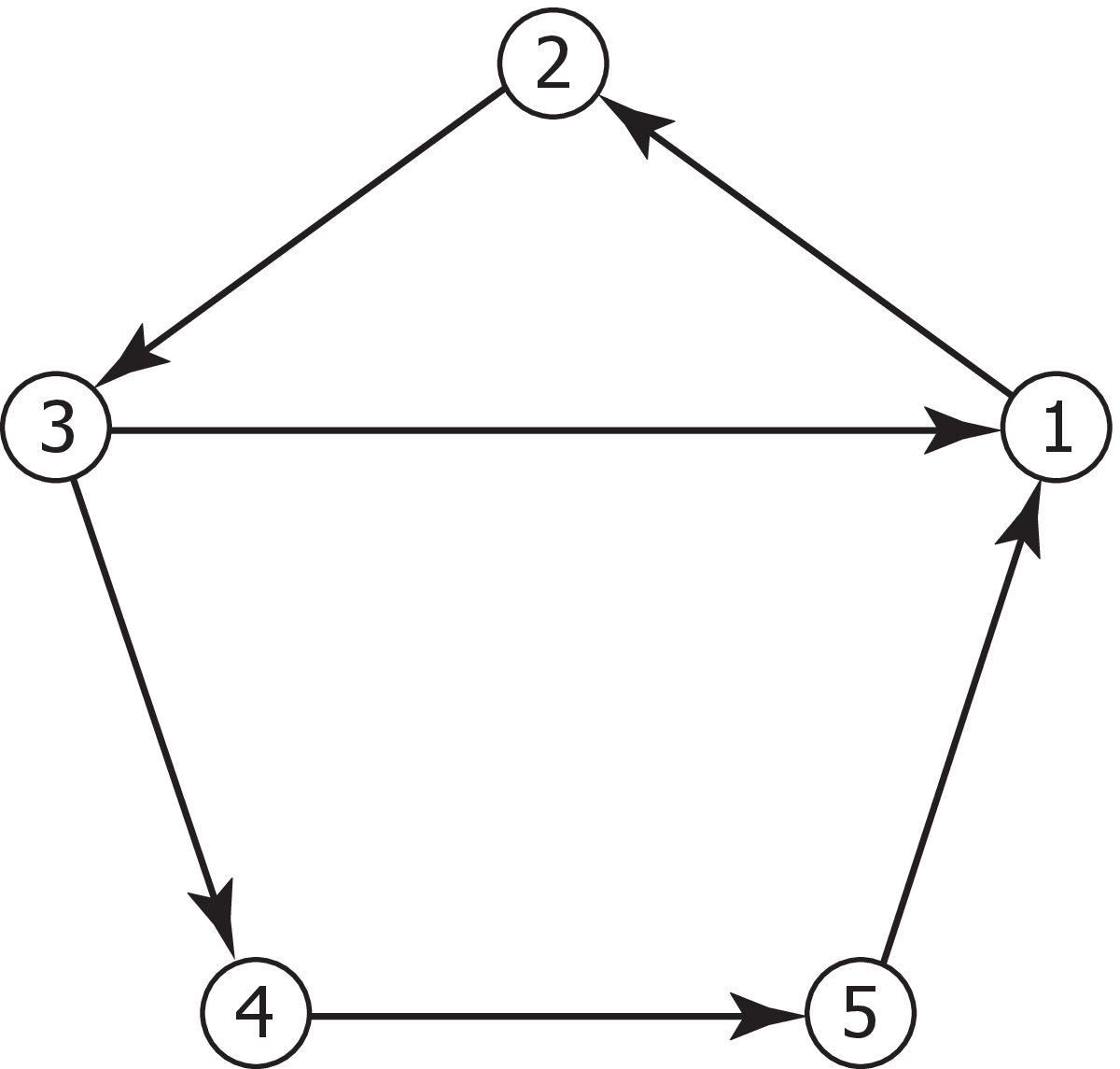}} \;\;&\;\;
       \subfigure[]{\includegraphics[width=.427\columnwidth]{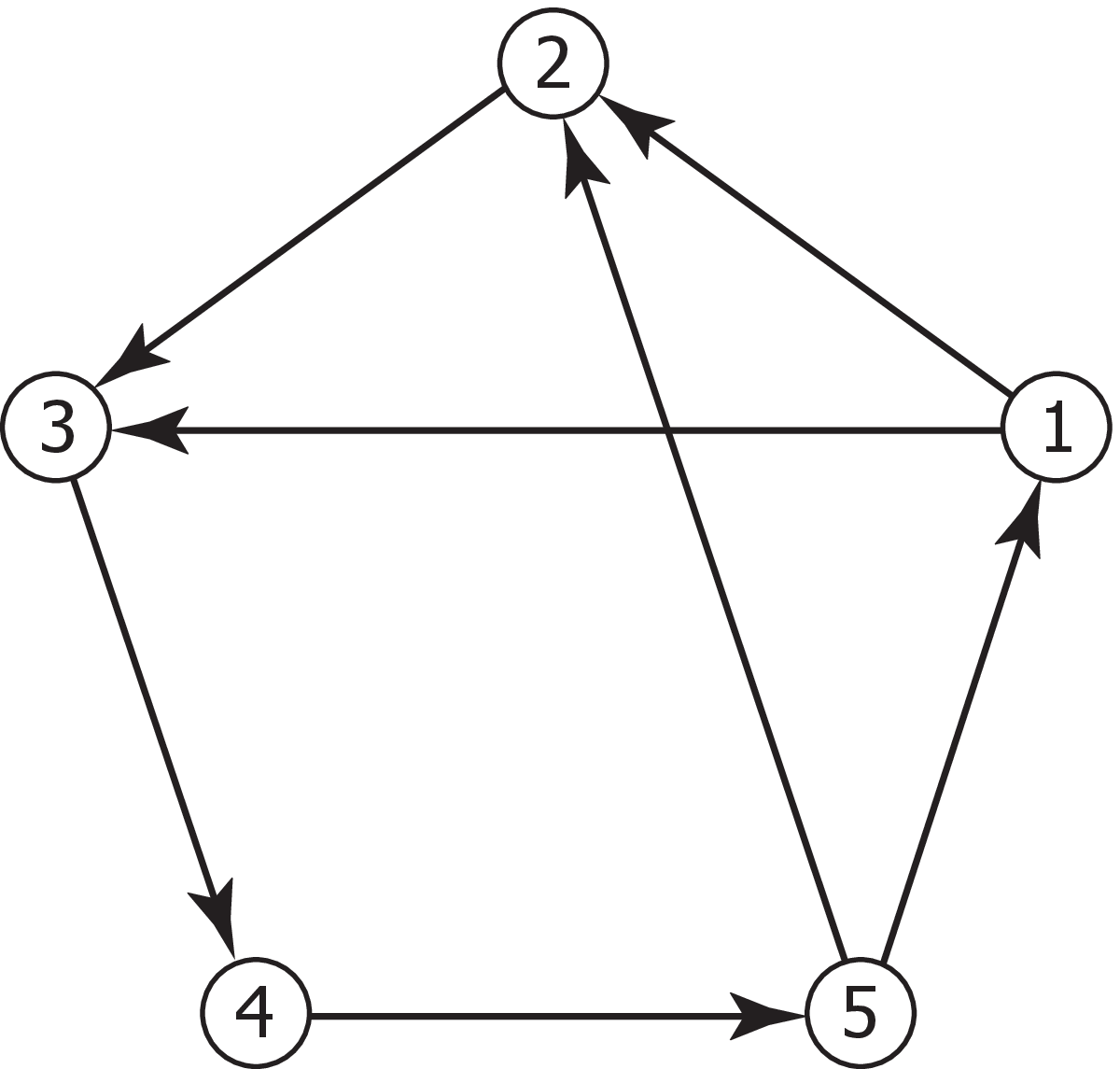}}
       \end{tabular}
       \vspace{-0.29cm}
       \caption{\emph{Example 2:} Variations on the directed cycle with five vertices.}\label{FIG:Example 2}
       \end{center}
\end{figure*}
Note that $\det(-\boldsymbol{\Delta}(\textsf{D}_n(s))) = (-1)^{n}\,(1 - s^n)$ 
and that the directed path and the directed cycle (for~$n$ even), are bipartite digraphs 
(cf. Prop.~\ref{Prop_Cycle} and Remark~\ref{Remark_bipart}).

For easiness of reference, the results found in this section are summarized in Table~\ref{Table2}.
\begin{remark}
Note that differently from Sect.~\ref{SEC:Prob}, since
$\boldsymbol{\Delta}(\mathcal{D}(s))$ is nonsymmetric,
it can also admit \emph{complex-conjugate eigenvalues} and the states
of system~(\ref{Eq_cons_defor_dir}) 
may experience stable steady-state oscillations (as we have seen
in Prop.~\ref{Prop_DirCyc} for $n$ odd). It is worth pointing out here
that this is not true, instead, for the protocol~(\ref{Eq_cons_dir}), 
which does not admit stable periodic solutions
(cf.~Prop.~3.10 in~\cite{MesbahiEg_book10}).~\hfill$\diamond$
\end{remark}
Since the study of the behavior of system~(\ref{Eq_cons_defor_dir}) for digraphs
$\mathcal{D}$ of arbitrary topology is nontrivial, we will focus here 
on few representative examples and try to deduce some criteria of
general validity.
\begin{example}\label{Example2}
Consider the four digraphs reported in~Fig.~\ref{FIG:Example 2}.
\begin{itemize}
\item With the digraph in Fig.~\ref{FIG:Example 2}(a), we have that
system~(\ref{Eq_cons_defor_dir}) is asymptotically stable for $s \in
(1/\cos(16\pi/5),\,1)$ (cf. Prop.~\ref{Prop_DirCyc}). For $s =
1/\cos(16\pi/5)$, system~(\ref{Eq_cons_defor_dir}) is marginally
stable: at steady-state, its states oscillate with the same frequency and amplitude, and the phases are evenly spaced.
For $s = 1$, average consensus is achieved (see the simulation results
in Sect.~\ref{Sect:DG}). 
\item With the digraph in Fig.~\ref{FIG:Example 2}(b),
system~(\ref{Eq_cons_defor_dir}) is asymptotically stable for $s \in
(-1.6889,\,1)$. For $s = -1.6889$, system~(\ref{Eq_cons_defor_dir}) is
marginally stable. At~steady-state, its states oscillate with the same
frequency but have different amplitudes: the phases are regularly
spaced. For \mbox{$s = 1$}, consensus (but not average consensus, since edge
$(1,\,3)$ breaks the balancedness of the original \mbox{five-vertex} directed
cycle), is achieved (see the simulation results in Sect.~\ref{Sect:DG}).
\item With the digraph in Fig.~\ref{FIG:Example 2}(c), we have the 
same qualitative behavior as with the digraph in 
Fig.~\ref{FIG:Example 2}(b). The only difference is represented by the stability threshold, that is now
$s = -1.9441$ instead of $s = -1.6889$.
\item With the digraph in Fig.~\ref{FIG:Example 2}(d),
system~(\ref{Eq_cons_defor_dir}) is asymptotically stable for $s \in
(-1.3326,\,1)$. For $s = -1.3326$, system~(\ref{Eq_cons_defor_dir}) is 
marginally stable and its \emph{non-oscillating} states asymptotically
converge to five different values. For $s = 1$, consensus (but, again, not average \mbox{consensus}) is achieved.
\end{itemize}
Note that we cannot leverage Prop.~\ref{Prop_stab_gen} to determine the stability interval 
of system~(\ref{Eq_cons_defor_dir}) when it admits stable periodic 
solutions (the threshold values 
for the digraphs in Figs.~\ref{FIG:Example 2}(b) and 
\ref{FIG:Example 2}(c), have been determined on a 
trial-and-error basis): however, Prop.~\ref{Prop_stab_gen} appears to provide the
correct stability intervals in all the other cases (e.g., for the
directed path in Fig.~\ref{FIG:Directed_graphs}(a), or for the digraph in 
Fig.~\ref{FIG:Example 2}(d)). 
The analytical determination of the stability
thresholds for general digraphs is the subject of 
\mbox{on-going} research. The directed graphs in Figs.~\ref{FIG:Example 2}(b) and \ref{FIG:Example 2}(c)
are particularly significative for showing the nontrivial connection
existing between the topology of the digraph $\mathcal{D}$ and 
the threshold values: in fact, here, a single edge-orientation change 
$($i.e., $(1,\,3)$ versus $(3,\,1))$, yields two remarkably different
thresholds: $s = -1.6889$ and $s = -1.9441$.~\hfill$\diamond$
\end{example}

\begin{figure*}[t!]
       \begin{center}
       \begin{tabular}{ccc}
       \hspace{-0.36cm}\subfigure[]{\includegraphics[width=.66\columnwidth]{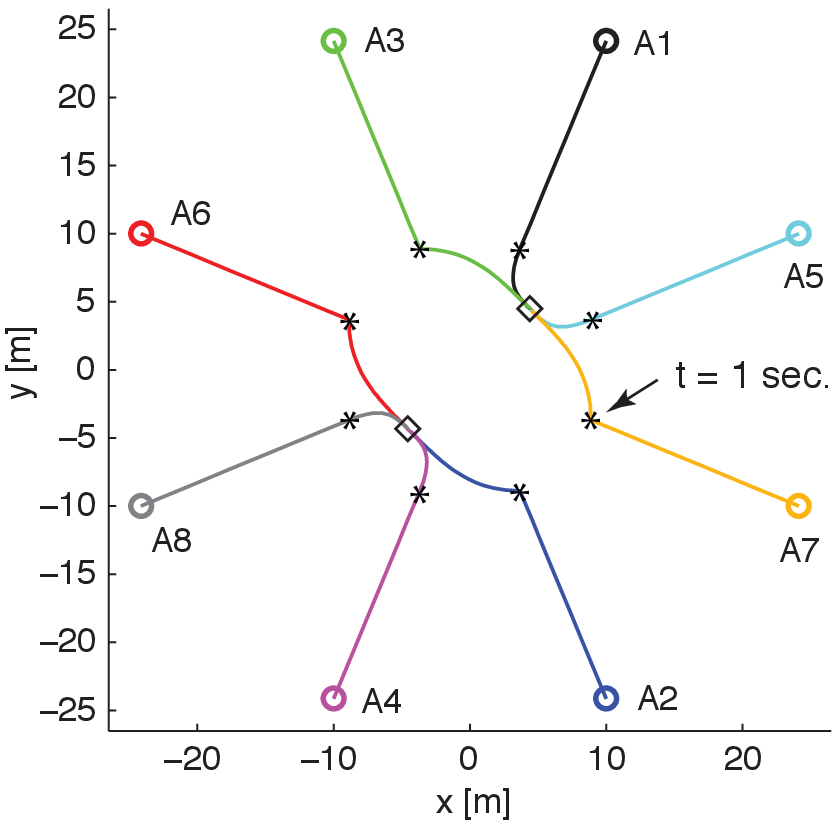}} \!\!&\!\!
       \psfrag{a}{\scriptsize{$p_{ix}$ [m]}}
       \psfrag{b}{\scriptsize{$p_{iy}$ [m]}}
       \subfigure[]{\includegraphics[width=.69\columnwidth]{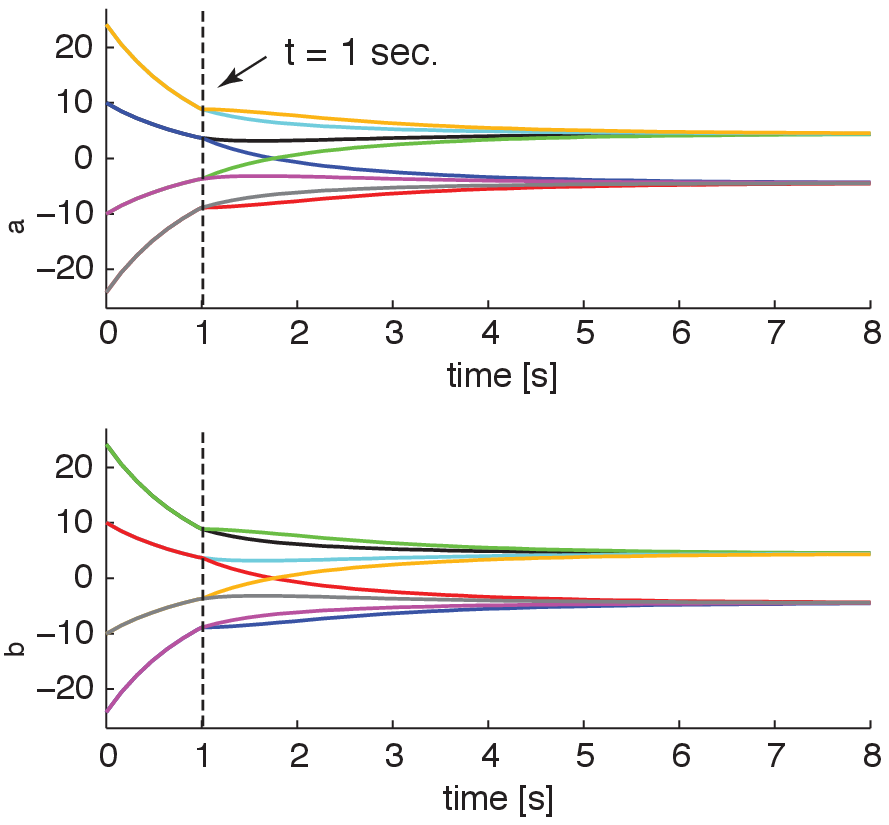}} &
       \psfrag{a}{\;\footnotesize{$s(t)$}}
       \subfigure[]{\includegraphics[width=.675\columnwidth]{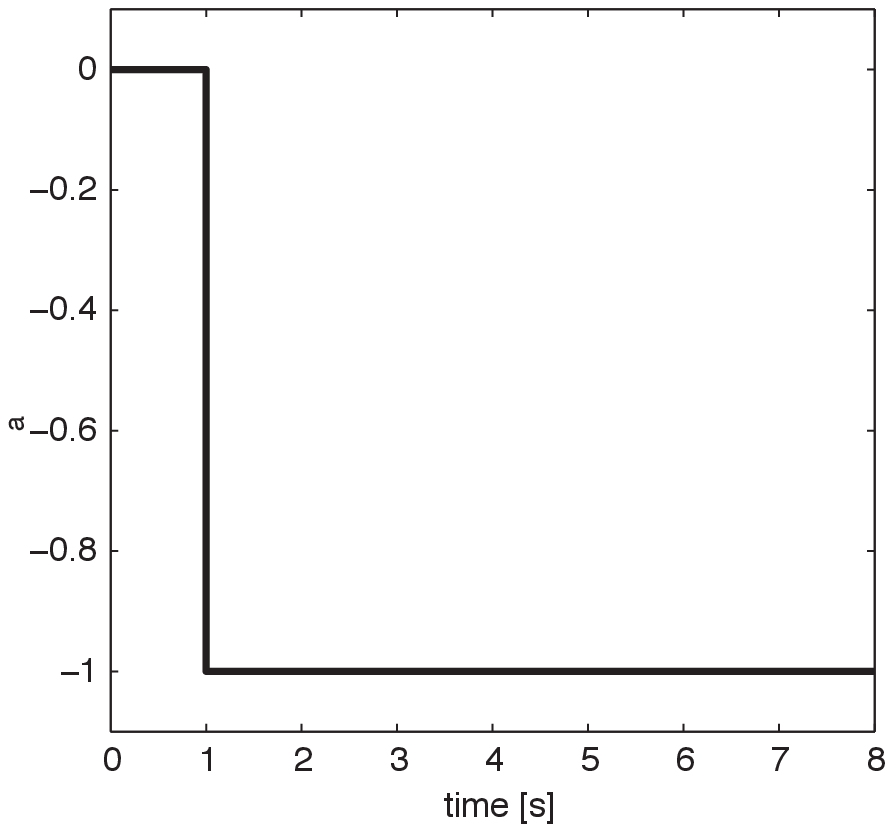}}
       \end{tabular}
       \vspace{-0.3cm}
       \caption{\emph{Simulation results $-$ undirected graph}: (a) Trajectory of the 8 vehicles:
       the communication topology is the cycle graph~$C_8$ (the
       initial position is marked with a circle and the final position
       with a diamond;  a star indicates the switching time); 
       (b)~Time evolution of the \mbox{$x$-, $y$-coordinates} of the vehicles
       (top~and bottom, respectively): the same color convention as in
       (a) is adopted here; (c) Time history of parameter~$s$.}\label{FIG:Simul}
       \end{center}
\end{figure*}
%

\section{Simulation results}\label{Sect:simul}
\vspace{-0.1cm}

Numerical simulations have been performed in order to
illustrate the theory presented in Sect.~\ref{SEC:Prob} and
Sect.~\ref{Sect:ext_Direct}. Sect.~\ref{Sect:UG} deals with the case
of undirected graphs, and Sect.~\ref{Sect:DG} with the case of directed graphs.

\subsection{Undirected communication graph}\label{Sect:UG}
\vspace{-0.1cm}

Consider a team of $n$ single-integrator agents,
$$
\dot{\vet{p}}_i(t) =\, \boldsymbol{\nu}_i(t),\;\; i \in \{1,\ldots,n\},
$$
where $\vet{p}_i(t) = [p_{ix}(t),\,p_{iy}(t)]^T \in \rr^2$ and
$\boldsymbol{\nu}_i(t) \in \rr^2$ denote respectively the position and
the input of vehicle~$i$ at time $t$. Let the control input
of agent $i$ be of the form,
\begin{equation}\label{Contr_prot}
\boldsymbol{\nu}_i(t) \,=\, (s^2 - 1)\,\vet{p}_i(t) \,+\,
s\!\!\!\sum_{j \,\in\, \mathcal{N}(i)} \!(\vet{p}_j(t) - s\,\vet{p}_i(t)),\vspace{-0.1cm}
\end{equation}
where $\mathcal{N}(i)$ denotes the set of vertices adjacent to vertex~$i$ 
in the communication graph\footnote{Note that we implicitly assume here that parameter $s$
is broadcast in real-time to all the agents by a supervisor, via a centralized transmitter.}. 
Then, the collective dynamics of the group of vehicles adopting control~(\ref{Contr_prot}),
can be written~as, 
$$
\dot{\vet{p}}(t) \,=\, (-\boldsymbol{\Delta}(s) \,\otimes\, \vet{I}_2)\,\vet{p}(t),
$$
where $\vet{p} = [\vet{p}_1^T,\,\ldots,\,\vet{p}_n^T]^T \in \rr^{2n}$
and ``$\otimes$'' denotes the Kronecker product.\\
Fig.~\ref{FIG:Simul}(a) shows the trajectory of 8~vehicles
implementing the control law~(\ref{Contr_prot}), 
when the communication topology is the cycle graph $C_8$
(the vehicles are initially on the vertices of a regular octagon
centered at the origin, and their position is marked with a circle in the figure).
For~the sake of illustration, in our simulation 
we selected the following switching signal
(see~Fig.~\ref{FIG:Simul}(c)): \vspace{-0.1cm}
$$
s(t) = \left\{
\begin{array}{lcl}
0\; &\,\text{for} & t \in [0,\, 1)~\text{sec}.,\vspace{0.09cm}\\
-1\; &\,\text{for} & t \in [1,\,8]~\text{sec}.
\end{array}
\right.
$$
The~time evolution of the \mbox{$x$-, $y$}-coordinates of the
agents is reported in Fig.~\ref{FIG:Simul}(b). As~it is evident
in Figs.~\ref{FIG:Simul}(a) and~\ref{FIG:Simul}(b), the vehicles first
converge towards the origin by maintaining equal interdistances, 
and then the even and odd agents cluster in two distinct groups (recall Prop.~\ref{Prop_Cycle}).
%
\begin{figure*}[t!]
       \begin{center}
       \begin{tabular}{ccc}
       \!\!\!\!\!\!\subfigure[]{\includegraphics[width=.68\columnwidth]{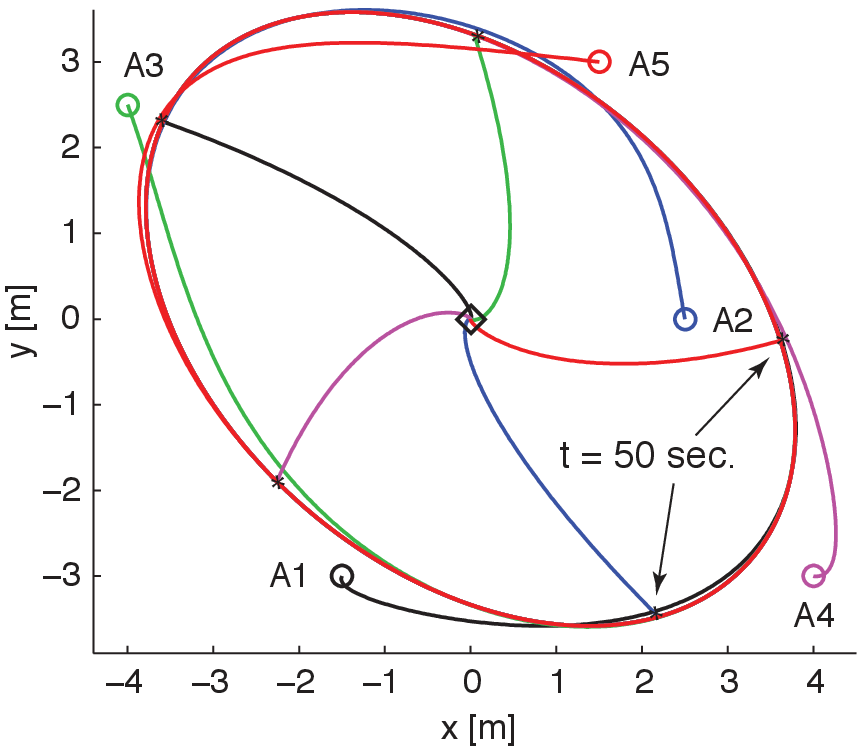}} \!\!&\!\!
       \psfrag{a}{\tiny{$p_{1x}$}}
       \psfrag{b}{\tiny{$p_{2x}$}}
       \psfrag{c}{\tiny{$p_{3x}$}}
       \psfrag{d}{\tiny{$p_{4x}$}}
       \psfrag{e}{\tiny{$p_{5x}$}}
       \psfrag{f}{\tiny{$p_{1y}$}}
       \psfrag{g}{\tiny{$p_{2y}$}}
       \psfrag{h}{\tiny{$p_{3y}$}}
       \psfrag{n}{\tiny{$p_{4y}$}}
       \psfrag{o}{\tiny{$p_{5y}$}}
       \subfigure[]{\includegraphics[width=.69\columnwidth]{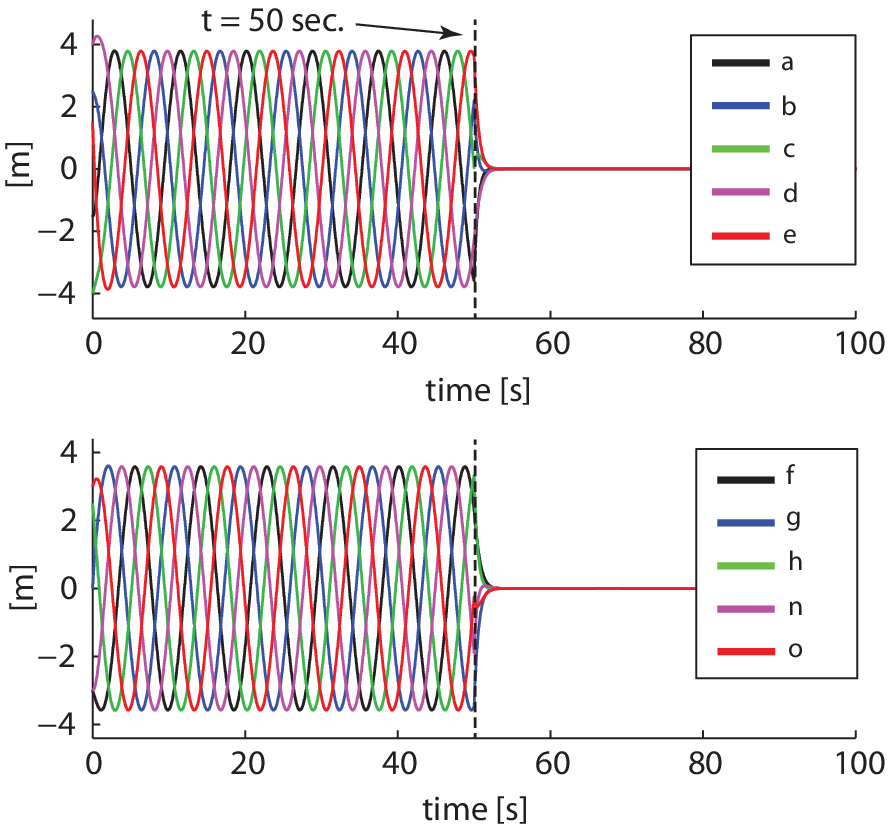}} &
       \psfrag{a}{$s(t)$\!\!}
       \psfrag{b}{\hspace{-0.3cm}\footnotesize{$1/\cos(16\pi/5)$}}
       \subfigure[]{\includegraphics[width=.67\columnwidth]{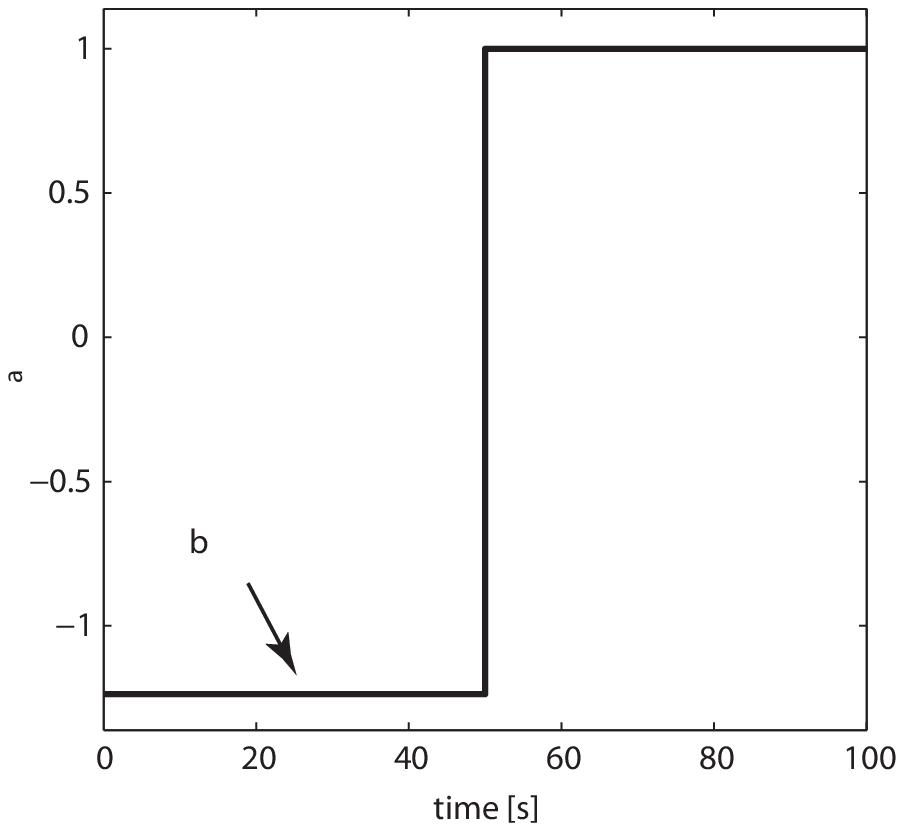}}\vspace{-0.15cm}\\
       \!\!\!\!\!\!\subfigure[]{\includegraphics[width=.68\columnwidth]{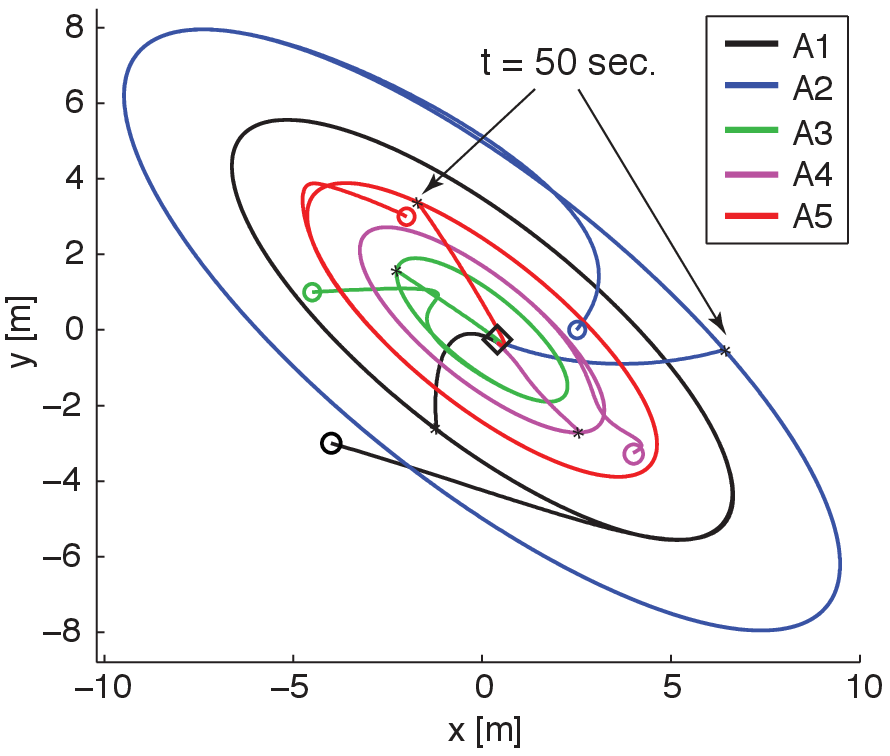}} \!\!&\!\!
       \psfrag{a}{\tiny{$p_{1x}$}}
       \psfrag{b}{\tiny{$p_{2x}$}}
       \psfrag{c}{\tiny{$p_{3x}$}}
       \psfrag{d}{\tiny{$p_{4x}$}}
       \psfrag{e}{\tiny{$p_{5x}$}}
       \psfrag{f}{\tiny{$p_{1y}$}}
       \psfrag{g}{\tiny{$p_{2y}$}}
       \psfrag{h}{\hspace{-0.01cm}\tiny{$p_{3y}$}}
       \psfrag{n}{\hspace{-0.01cm}\tiny{$p_{4y}$}}
       \psfrag{o}{\hspace{-0.01cm}\tiny{$p_{5y}$}}
       \subfigure[]{\includegraphics[width=.69\columnwidth]{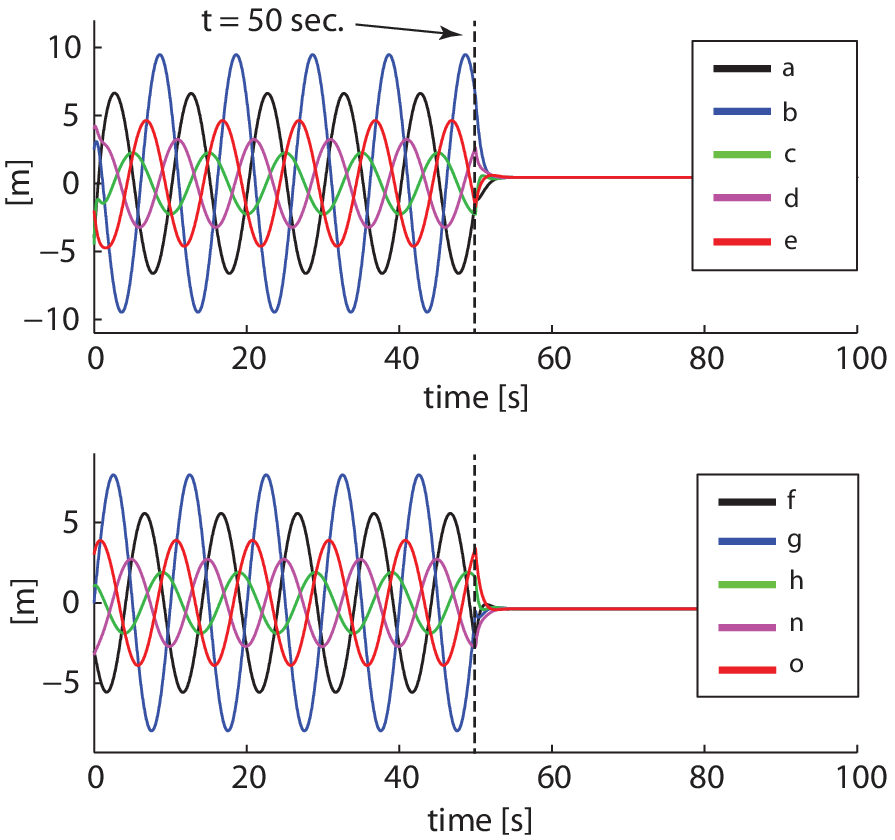}} &
       \psfrag{a}{$s(t)$\!\!}
       \psfrag{b}{\!\!\footnotesize{$-1.6889$}}
       \subfigure[]{\includegraphics[width=.67\columnwidth]{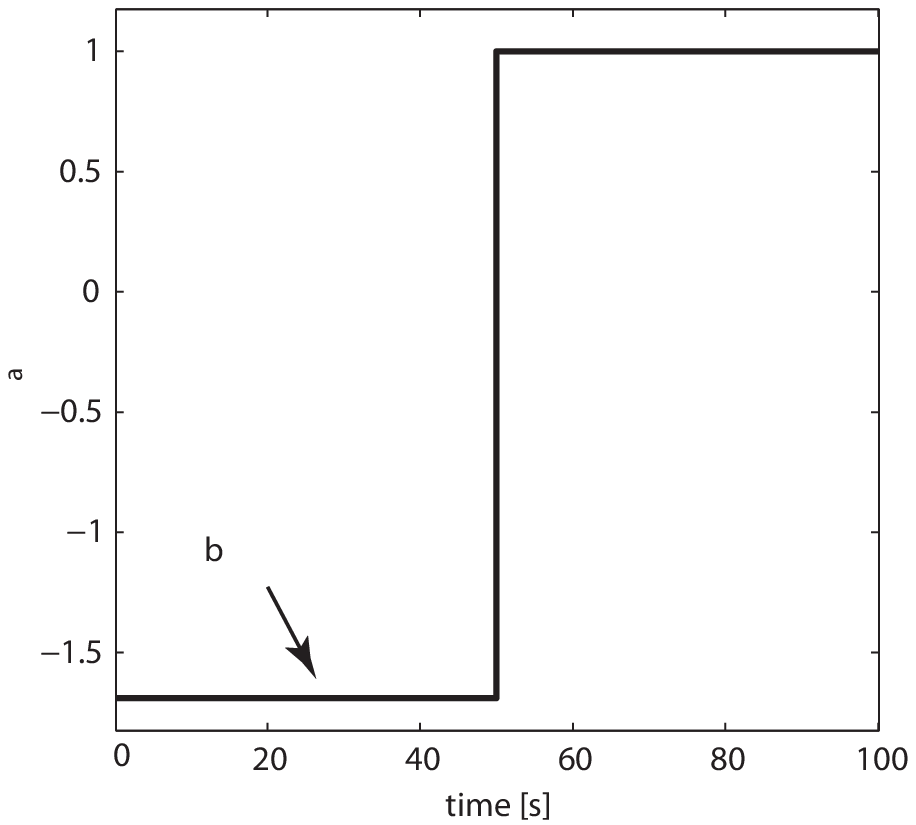}}
       \end{tabular}
       \vspace{-0.4cm}
       \caption{\emph{Simulation results $-$ directed graph}: (a)
         Trajectory of the 5 vehicles: the communication topology is
       the digraph in Fig.~\ref{FIG:Example 2}(a), (the initial
       position is marked with a circle and the final position with a
       diamond; a star indicates the switching time); (b)~Time evolution of the \mbox{$x$-,
         $y$-coordinates} of the vehicles (top~and bottom, respectively); (c) Time history of
       parameter~$s$; (d)~Trajectory of the 5 vehicles: the
       communication topology is the digraph in Fig.~\ref{FIG:Example
         2}(b); (e)~Time evolution of the \mbox{$x$-, $y$-coordinates} of the
       vehicles; (f) Time history of parameter~$s$.}\label{FIG:Simul2}
       \end{center}
\end{figure*}

\vspace{-0.08cm}
\subsection{Directed communication graph}\label{Sect:DG}
\vspace{-0.1cm}
A scenario similar to that described in Sect.~\ref{Sect:UG} is considered in Fig.~\ref{FIG:Simul2}.
In this case, 5 single-integrator agents implement the control law~(\ref{Contr_prot}) and 
communicate using a directed graph. This leads to an overall closed-loop system of the form,
\begin{equation*}\label{Close_loop_dir}
\dot{\vet{p}}(t) \,=\, (-\boldsymbol{\Delta}(\mathcal{D}(s)) \,\otimes\, \vet{I}_2)\,\vet{p}(t).
\end{equation*}
The results in Figs.~\ref{FIG:Simul2}(a)-(c), are relative to
the digraph in Fig.~\ref{FIG:Example 2}(a), i.e., $\textsf{D}_5$. Fig.~\ref{FIG:Simul2}(a) shows the trajectory of the
5 vehicles starting from the position $\vet{p}(0) = [-1.5,\, -3,\, 2.5,\, 0,\, -4,\, 2.5,\, 4,\, -3,\,1.5,\, 3]^T$.
As~shown in Fig.~\ref{FIG:Simul2}(d),\vspace{-0.1cm}
$$
s(t) = \left\{
\begin{array}{lll}
1/\cos(16\pi/5)\; &\text{for} & t \in [0,\, 50)~\text{sec}.,\vspace{0.08cm}\\
1\; &\text{for} & t \in [50,\, 100]~\text{sec}.\vspace{-0.1cm}
\end{array}
\right.
$$
Fig.~\ref{FIG:Simul2}(b) displays the time evolution of the \mbox{$x$-, $y$}-coordinates of the
agents. From Figs.~\ref{FIG:Simul2}(a) and~\ref{FIG:Simul2}(b), we can see that the vehicles first
move counterclockwise along a common elliptical trajectory with frequency $f(5) =
\frac{1}{2\pi}\tan(16\pi/5)$ and phases $\phi_i(5) =
\frac{2\pi(i-1)}{5\tan(16\pi/5)}$, $i \in \{1,\ldots,5\}$, 
and then rendezvous at the point,\vspace{-0.1cm}
$$
\Big(\,\frac{1}{5}\sum_{i=1}^5 p_{ix}(50),\;\;\frac{1}{5}\sum_{i=1}^5 p_{iy}(50)\Big),
$$ 
i.e., they achieve average consensus (recall Prop.~\ref{Prop_DirCyc}).\\
Finally, the results in Figs.~\ref{FIG:Simul2}(d)-(f), are relative to
the digraph in Fig.~\ref{FIG:Example 2}(b). Fig.~\ref{FIG:Simul2}(d) shows the trajectory of the
agents with $\vet{p}(0) \!=\! [-4, -3, 2.5, 0, -4.5, 1, 4, \!-3.25, \!-2, 3]^T$\!.
In this case (see Fig.~\ref{FIG:Simul2}(f)), parameter $s$ evolves
\mbox{according to:}\vspace{-0.1cm} 
$$
s(t) = \left\{
\begin{array}{lll}
-1.6889\; &\text{for} & t \in [0,\, 50)~\text{sec}.,\vspace{0.08cm}\\
1\; &\text{for} & t \in [50,\, 100]~\text{sec}.
\end{array}
\right.
$$
From Figs.~\ref{FIG:Simul2}(d) and~\ref{FIG:Simul2}(e), we see
that the vehicles first move counterclockwise on five closed orbits, and then
rendezvous at a single point (recall item 2 in \mbox{Example~\ref{Example2}}).\\
It is worth observing here that the two simple control instances described in
this subsection, look promising building blocks for more sophisticated multi-agent 
tasks, such as, e.g., for 
containment control~\cite{JiFeEgBu_TAC08} or cooperative patrolling~\cite{PasqualettiFrBu_TRO12}.

\vspace{-0.25cm}
\section{Conclusions and future work}\label{Sect:concl}
\vspace{-0.25cm}
In this paper we have presented a generalization of the standard consensus protocol,
called \emph{deformed consensus protocol}, and we have analyzed its stability properties
in terms of the real parameter $s$ for some special families of undirected and directed graphs.
Preliminary results for arbitrary graph topologies are also provided: however, some work still needs
to be done in order to precisely characterize in graph-theoretical
terms, the variegated behavior of the deformed consensus protocol
(a~glimpse of such a richness of behaviors is provided
by Examples~\ref{Example} and~\ref{Example2}). The proposed theory has been illustrated
via \mbox{extensive} numerical simulations and examples.\vspace{0.05cm}\\
In future works, we aim at studying the properties of the deformed consensus protocol when the
(weighted) communication graph is not fixed but changes over time, 
at establishing a link with the existing cluster synchronization
and group consensus literature~\cite{YuWa_SCL10,XiaCa_AUTO11},
and at investigating other ``parametric'' Laplacian matrices besides
the deformed Laplacian (a first step towards this direction has 
been recently done in~\cite{Morbidi_TAC13}). 
%
%
%
%
%
%
\bibliographystyle{plain} 
\bibliography{biblio_ext}
\end{document}